\documentclass[preprint,aps,superscriptaddress,nofootinbib,showkeys,10pt]{revtex4}
\pdfoutput=1
\usepackage{graphicx}
\usepackage{amsmath}
\usepackage{amsfonts}
\usepackage{mathtools}
\usepackage{placeins}
\usepackage{xparse}
\usepackage{physics}
\usepackage{subfigure}
\usepackage{slashed}
\usepackage{multirow}
\usepackage{csquotes}
\usepackage{lscape}
\usepackage{float}
\usepackage{titlesec}
\usepackage{amssymb}
\usepackage{xcolor, soul}
\usepackage{epstopdf}
\usepackage{hyperref}
\hypersetup{
     colorlinks   = true,
     citecolor    = red,
     linkcolor    = blue,
     urlcolor     = blue,
}
\definecolor{deeppink}{rgb}{0.9, 0.17, 0.31}
\def\beq{\begin{equation}}
\def\eeq{\end{equation}}
\def\bea{\begin{eqnarray}}
\def\eea{\end{eqnarray}}
\def\be{\begin{equation}}
\def\ee{\end{equation}}

\def\nno{\nonumber}
\def\bse{\begin{subequations}}
\def\ese{\end{subequations}}

\graphicspath{{./figs/}}
\begin{document}

\title{Squeezing, Chaos and Thermalization in Periodically Driven Quantum Systems:\\ The Case of Bosonic Preheating \\
}

\author{Ayan Chakraborty}%
\email{chakrabo@iitg.ac.in}
\author{Debaprasad Maity}
\email{debu@iitg.ac.in}
\affiliation{%
	Department of Physics, Indian Institute of Technology Guwahati.\\
	Guwahati, Assam 781039, India 
}%

\date{\today}
\begin{abstract}
The phenomena of Squeezing and chaos have recently been studied in the context of inflation. We apply this formalism in the post-inflationary preheating phase. During this phase, inflaton field undergoes quasi-periodic oscillation, which acts as a driving force for the resonant growth of quantum fluctuation or particle production. Furthermore, the quantum state of the fluctuations is known to have evolved into a squeezed state. In this submission, we explore the underlying connection between the resonant growth, squeezing, and chaos by computing the Out of Time Order Correlator (OTOC) of phase space variables and establishing a relation among the Lyapunov, Floquet exponents, and squeezing parameters. For our study, we consider observationally favored $\alpha$-attractor E-model of inflaton which is coupled with the bosonic field. After the production, the system of produced bosonic fluctuations/particles from the inflaton is supposed to thermalize, and that is believed to have an intriguing connection to the nature of chaos of the system under perturbation. 
We conjecture a relation between the thermalization temperature $({\bar T}_{\rm SS})$ of the system and quantum squeezing, which is further shown to be consistent with the well-known Rayleigh-Jeans formula for the temperature symbolized as ${\bar T}_{\rm RJ}$, and that is ${\bar T}_{\rm SS} \simeq {\bar T}_{\rm RJ}$. Finally, we show that the system temperature is in accord with the well-known lower bound on the temperature of a chaotic system proposed by Maldacena-Shenker-Stanford (MSS). 
\end{abstract}
\maketitle
\section{\textbf{ Introduction}}\label{sec1}
Post-inflationary reheating is an important event in the early universe that is responsible for the thermalized universe we see today. Therefore, a theoretical understanding of this phase would be of paramount interest. Inflationary quantum fluctuations are responsible for producing curvature fluctuations, and their subsequent generation of structures is inevitably related to the process of transferring energy from inflaton to matters.  
In spite of having 
significant efforts  \cite{Guth:1980zm,Senatore:2016aui,Linde:1981mu,Albrecht:1982wi,Lemoine:2008zz,Baumann:2018muz,Piattella:2018hvi,Kofman:1994rk,Kofman:1997yn,Greene:1997fu,Greene:2000ew,Lozanov:2019jxc,Tsujikawa:1999me,Shtanov:1994ce,Kaiser:1995fb,Amin:2014eta,Podolsky:2002qv,Palma:2000md,ABBOTT198229,Dolgov:1982th,Garcia:2020wiy,Garcia:2020eof,Haque:2020zco,Haque:2022kez,Haque:2023yra}, a proper theoretical understanding of this phase is still lacking particularly due to its complex non-linear evolution, and subsequent thermalization.    

After a quasi-exponential accelerated phase of expansion, known as inflation \cite{Guth:1980zm,Senatore:2016aui,Linde:1981mu,Albrecht:1982wi,Lemoine:2008zz,Baumann:2018muz,Piattella:2018hvi},  
the universe is left with a super cold state of vanishing matter, entropy density, and a homogeneous inflaton field oscillating around its potential minima. 
When coupled with such oscillating inflaton, any fundamental field will experience a periodic driving force and have quantum mechanical production. Such production, therefore, leads to the transfer of energy from inflaton to standard model fields successfully populating the universe.
 Typically such a transferring process involves multiple stages of evaluations. Depending upon the coupling parameter, the initial stage may be non-perturbative \cite{Kofman:1994rk,Kofman:1997yn,Greene:1997fu,Greene:2000ew,Lozanov:2019jxc,Tsujikawa:1999me,Shtanov:1994ce,Kaiser:1995fb,Amin:2014eta,Podolsky:2002qv,Palma:2000md} followed by perturbative production, or the process could be entirely perturbative \cite{ABBOTT198229,Dolgov:1982th,Garcia:2020wiy,Garcia:2020eof,Haque:2020zco,Haque:2022kez,Haque:2023yra,Chakraborty:2023ocr}. However, apart from non-linear nature of production, it is the thermalization of those produced particles in the expanding background which makes the whole process extremely complex and theoretically challenging. Studies of perturbative production and its thermalization have been discussed in the literature \cite{Son:1996uv,McDonald:1999hd,Harigaya:2013vwa,Mukaida:2015ria} taking a phenomenological approach with very little foundational progress. Considering the lattice simulation, studies have also been performed for non-perturbative production and their subsequent thermalization \cite{Podolsky:2005bw,Felder:2000hr,Desroche:2005yt,Micha:2004bv,Maity:2018qhi}. However, to the best of our knowledge, a proper underlying mathematical framework to understand this thermalization process in the context of reheating is far from complete. It is in this realm, that we initiate formulating a theoretical framework that we believe to play an instrumental role in understanding such a phase. Apart from its theoretical motivation, it is important to stress that the production and thermalization during reheating may leave an indelible imprint on different cosmological observable related to Cosmic Microwave Background (CMB), baryogenesis, dark matter, and even small-scale structures in the universe. 

In this submission, we particularly focus on the stage called preheating when particle production occurs with parametric resonance as a primary mechanism, and hence would naturally last for a very brief initial period of the entire reheating process. The formalism we adopt is based on the recent development \cite{Martin:2015qta,Martin:2019wta,Haque:2020pmp,Bhattacharyya:2020kgu,Choudhury:2020yaa,Bhargava:2020fhl} on the application of various intriguing concepts of quantum information theory in the realm of cosmology particularly centered around one of the fundamental questions in theoretical cosmology as to how the quantum correlation of fluctuations generated during inflation is dynamically decohered into post-inflationary classicalized correlation, and what would be the appropriate quantifier of their quantumness \cite{Martin:2015qta,Martin:2022kph}.  
During inflation, the primordial perturbation generically evolves into a squeezed state \cite{Albrecht:1992kf,Grishchuk:1990bj}. 
Utilizing this squeezed state and to decode the quantumness of the inflationary correlations, a number of proposals have been put forth based on quantum information tools in the recent past namely, observing the Bell violation \cite{Martin:2019wta,Kanno:2017teu,Kanno:2017dci}, quantum discord \cite{Martin:2015qta,Kanno:2016gas,Ollivier:2001fdq,Henderson_2001}, quantum stokes parameter \cite{Maity:2021zng}.
In the quantum field theory framework, on the other hand, the connection between squeezing and chaos has been known for quite some time \cite{Alekseev_1998,PhysRevE.79.046220}.
The growing interest in quantum information theory further unveiled the idea of OTOC, complexity as an important diagnostic tool to characterize the chaos first observed in the context of information scrambling by Black holes \cite{Sekino:2008he,Hayden:2007cs,Maldacena:2015waa}. Utilizing those tools quantum chaotic growth \cite{Haque:2020pmp,Bhattacharyya:2020kgu} of primordial perturbation has also been quantified through diagnostics such as OTOC \cite{Hashimoto:2017oit,Rozenbaum:2016mmv,Das:2021qsd}, complexity \cite{Ali:2019zcj,Yang:2019iav,Balasubramanian:2019wgd}, and are shown to have signatures in the correlations \cite{Park:2022xyf}.

In this work, as described above, we shall make use of those different tools and explore interconnection among resonance, squeezing, and chaos during the preheating phase. We particularly show that the production under the background of periodic inflaton field during preheating is an ideal quantum chaotic system, where the OTOC of phase space operators exhibits an exponential growth $ \sim e^{2\lambda t}$ in time. Where, $\lambda$, the Lyapunov exponent characterizes the chaotic nature of the system under consideration. 
A few earlier isolated attempts in this direction can be found in \cite{Joras:2001yh,Jin:2004bf}, where the chaotic nature of the preheating phase has been indicated without detailed characterization. We performed a detailed study on this by calculating OTOC as a diagnostic tool for a class of $\alpha-$attractor E-model of inflaton and different coupling with daughter fields. 
We extend our exploration further and shed light on how to quantify the possible thermalization temperature of the system in the parlance of squeezing formalism. From the perspective of a particular observer, the squeezed state is found to be thermal. Utilizing this observation, we arrive at an interesting relation between the squeezing and temperature of the system under consideration. At this juncture, we must remind the reader of a deep connection between chaos and thermalization as conjectured by Maldacena-Shenker-Stanford (MSS)\cite{Maldacena:2015waa} and this states that for any thermal quantum chaotic system, the characteristic Lyapunov exponent ($\lambda$) is constrained by its thermalization temperature $\lambda \leq 2\pi {k_B T}/{\hbar}$. Therefore, for any thermal quantum chaotic system, the temperature is set to have an upper bound fixed by the characteristic exponent $\lambda$. The temperature defined from the squeezing is indeed found to satisfy such an upper bound. 
 
 The order of construction of this paper is as follows. In Section \ref{sec2}, we first give the model of inflaton background specifying different potential model parameters. We consider $\alpha$-attractor E-model which is one of the observationally favored models in the context of early inflation, and reheating. In Section \ref{sec3}, we construct our model in a two-mode squeezed state language. We derive three dynamical equations of squeezing parameter($r_k$), squeezing angle($\varphi_)$, rotation angle($\theta_k$), and the dynamics of these parameters signify the evolution of the quantum state of the produced fluctuation. We also obtain an equivalent expression of occupation number density in terms of squeezing parameter. In Section \ref{sec4}, we introduce the concept of OTOC, one of the diagnostics of quantum chaos, and we construct the quantum counterpart of classical \textit{symplectic matrix} taking different combinations of field and its conjugate momenta. Finally, we shall calculate the OTOC of the system from the dominant eigenvalue of the matrix. In Section \ref{sec5}, we discuss different inflationary parameters along with the dynamics of inflaton as a periodic driving source for three different models. The very dynamics of this driving force make the underlying chaotic nature and thermalization process distinguishable from one background model to another. In Section \ref{sec6}, we introduce two different interactions between the produced fluctuation and background. Here we consider two well-known three-leg and four-leg type interactions in the QFT framework. In the post-inflationary reheating phase, the production of elementary particles is widely studied by taking two different effects, perturbative and non-perturbative effects into account.  Here we estimate a minimum bound on coupling strength above which fluctuation will be resonantly produced in a dynamical background. From Section \ref{sec8}, we calculate OTOC in two-mode squeezed state formalism for four-leg type interaction, and emphasize distinct features of its dynamics for three different background driving sources. In Section \ref{sec9}, we repeat exactly the same analysis of Section \ref{sec8} for three-leg type interaction. In Section \ref{sec10}, we derive a relation between resonant growth index(\textit{Floquet exponent}) and chaotic growth index(\textit{Lyapunov exponent}) and also study the behavior of the OTOC spectrum. In Section \ref{sec12}, we wish to have a semi-classical visualization of the quantum chaotic system by generating Poincar$\acute{\text{e}}$ section. Furthermore, we give an approximate estimate of the thermalized temperature of the system which is calculated from the squeezed state and Rayleigh-Jeans spectrum. We also check the consistency of these outcomes with MSS conjecture. Finally in Section \ref{sec13}, we conclude the paper stating some possible future direction of this work.
\section{Inflation and general model of Preheating}\label{sec2}

\subsection{ Description of background: Models and equations }
We will construct our work on the basis of the consideration of a homogeneous, isotropic universe being described by the (observationally favored, spatially flat) Freedmann-Lametre-Robertson-Walker (FLRW) metric,
\begin{equation}\label{2.1}
    ds^2=-dt^2+a^2(t)d\vec{x}^2 ,
\end{equation}
where $a(t)$ is scale factor in cosmic time $t$.
At the end of quasi-exponential accelerated era, called inflation, inflaton field starts oscillating around its potential minimum. Because of the expansion of the background, the oscillation amplitude falls with time, resulting in a gradual decrease of Hubble parameter \big($H={\dot a(t)}/{a(t)}\big)$. The dynamics governing the background evolution are well-known Friedmann equations,
\begin{equation}\label{2.2}
    \ddot \phi +3H\dot \phi+\pdv{ V(\phi)}{\phi}=0, \quad H^2=\frac{1}{3M_{\rm pl}^2}\bigg(\frac{1}{2}\dot \phi^2+V(\phi)\bigg) ,
\end{equation} 
where $M_{\rm pl}={1}/{\sqrt{8\pi G}}\approx 2.435\cross 10^{18} \text{GeV}$ is the reduced $Planck$ mass, and $V(\phi)$ is the inflaton potential. Solutions of the Eq.(\ref{2.2}), subject to slow roll initial conditions constrained by CMB observation ($n_s,r$), give us the evolution of the background and Hubble scale after inflation. In order to proceed we will consider a class of observationally favored $\alpha$-attractor E-model with the potential \cite{Martin:2013tda,Martin:2013nzq,Cook:2015vqa,Drewes:2017fmn},  
\begin{equation}\label{2.3}
V(\phi)=\Lambda^4\Big(1-e^{-\sqrt{\frac{2}{3\alpha}}.\frac{\phi}{M_{\rm pl}}}\Big)^{2n} .
\end{equation}
By varying the exponent \enquote{$n$} we can achieve different power law forms of the potential namely, quadratic model(for $n=1$), quartic model (for $n=2$), and so on. 
The amplitude of the potential \enquote{$\Lambda$}, which measures the energy content in the inflaton, is constrained by the CMB measurement, and is related to the scalar spectral index $n_s$, the amplitude of the inflaton fluctuation measured as CMB normalization $A_s = 3\times 10^{-9}$ and tensor to scalar ratio $r$. The model is favored by Planck 2018 observation \cite{Planck:2018vyg}, where $n_s=0.9649\pm 0.0042$ at $68\%$ CL and $95\%$ CL upper limit on tensor-to-scalar ratio $r_{0.002}$ is obtained as $r_{0.002}<0.056$ combining with  the BICEP2/Keck Array BK15 data. We will be using these observations throughout. 
The tensor-to-scalar ratio $r$ can be analytically expressed as,
\[ r= \frac{192\alpha n^2(1-n_s)^2}{\Big[4n+\sqrt{16n^2+24\alpha n(1-n_s)(1+n)}\Big]^2}\]
derived in \cite{Drewes:2017fmn}. 
Another parameter $(\alpha)$ determines the shape of the potential. The energy scale of inflation related to the parameter $\Lambda$ can be analytically expressed in terms of CMB parameters as \cite{Drewes:2017fmn}.

\begin{equation}\label{2.4}
      \Lambda= M_{\rm pl}\Big(\frac{3\pi^2rA_s}{2}\Big)^{\frac{1}{4}}\Bigg[\frac{2n(1+2n)+\sqrt{4n^2+6\alpha(1+n)(1-n_s)}}{4n(1+n)}\Bigg]^{\frac{n}{2}}
   \end{equation}
    In our work, we will obtain different backgrounds specifying different values of the exponent \enquote{$n$} in the given potential model (\ref{2.3}).

   \subsection{Preheating Stage After Inflation}
Just after the end of inflation, inflaton field amplitude is expected to be high($\phi \sim M_{\rm pl}$ ) and consequently, the energy stored in this field is also very high. In this stage, instead of behaving like an individual inflaton particle(quanta), the entire field behaves like a \textit{coherently oscillating inflaton field}. This behavior can also be thought of as a condensate formed by the superposition of all zero momentum inflaton particles and coherently oscillating in the expanding background. Therefore, any other fields directly/indirectly coupled with this coherently oscillating inflaton will experience a periodic driving force, and if the coupling strength is appropriate, the field amplitude will be resonantly enhanced. It is this coupling regime where we will do the detailed theoretical exploration of the phenomena of particle production, resonance, squeezing, and chaos of the system that emerge in the early universe cosmology. Such phenomena is known as {\it preheating}, which has been studied quite extensively in the literature (See earlier references \cite{Kofman:1994rk,Kofman:1997yn}). However, in this paper, we will redo this analysis from the perspective of quantum squeezing phenomena and its connection with chaos and thermalization. The span of this era is generically very short proportional to inverse decay width and happens within a few e-folds. In this short span explosive particle production takes place through \textit{parametric resonance}, hence the energy is transferred very rapidly from the inflaton field to the daughter fields. In this paper, we consider the produced daughter fields to be bosonic in nature. 
This idea of explosive production through parametric resonance was first put forward and thoroughly studied in the phenomenal work by Linde, Kofman and Starobinsky \cite{Kofman:1997yn}.

To proceed toward the quantitative discussion of this production process, we tart with a general Lagrangian containing different interaction terms corresponding to different fields. The form of the general Lagrangian for inflaton $(\phi)$, daughter scalar ($\chi$), and fermion $(\psi)$ is taken to be, 
\begin{equation}\label{2.5}
    L=\frac{1}{2}\partial_{\mu}\phi\partial^{\mu}\phi-V(\phi)+\frac{1}{2}\partial_{\mu}\chi\partial^{\mu}\chi-\frac{1}{2}m_{\chi}^2\chi^2 +\Bar\psi(i.\gamma^{\mu}\partial_{\mu}-m_{\psi}\psi-\frac{1}{2} F(\phi, R) \chi^2-h\bar\psi\psi\phi .
\end{equation}
Where, $V(\phi)$ is inflaton potential, $m_{\chi}$ is the bare mass of the produced scalar particle. $F(\phi, R)$ is the generic coupling term involving the daughter field and Ricci scalar $(R)$ and background inflaton.. The last term is the coupling between fermionic and background inflaton with the dimensionless coupling constant \enquote{$h$}. Considering different interaction terms, we can describe the explosive production of different elementary particles from a time-dependent background. In our present work, we will mainly concentrate on bosonic preheating. Fermionic one we left for our future work. Our aim is to study the phenomena of preheating in squeezed state formalism and its theoretical implications in understanding non-equilibrium phenomena such as chaos, thermalization.  
\section{Squeezed state formalism }\label{sec3}
In this section, we will build up the analytical structure of the model in a two-mode squeezed state framework which will be the mainstay of our numerical studies later (see earlier attempt \cite{Hirai2000SqueezePI}). Here, we mainly concentrate on scalar modes which are parametrically excited due to inflaton considering only bosonic interaction Lagrangian ,
\begin{equation}\label{3.1}
    \mathcal{L}= -\frac{1}{2}\sqrt{-g}\Big(\partial_{\mu}\chi \partial^{\mu}\chi+ m_{\chi}^2\chi^2+ F(\phi) \chi^2\Big) .
\end{equation}
Where $\sqrt{-g}=a^3(t)$, $m_{\chi}$ is the bare mass of produced particles, and $F(\phi)$ is a generic coupling function of the background inflaton ($\phi$) and specific form of this function will be discussed later in Section \ref{sec6}. 
We ignore the Ricci curvature coupling for simplicity. 
Expressing the scalar field ($\chi$) in terms of Fourier modes , 
\begin{equation}\label{3.2}
     \chi(t,\vec{x})= \int\frac{d^3\vec{k}}{(2\pi)^3} \chi_k(t) e^{i\vec{k}.\vec{x}}, 
 \end{equation}
 the Lagrangian becomes,
  \begin{equation}\label{3.3}
      L=\frac{1}{2}\int\frac{d^3\vec{k}}{(2\pi)^3}\Big[a^3\dot\chi_{\vec{k}}\dot\chi_{-\vec{k}}-a.k^2\chi_{\vec{k}}\chi_{-\vec{k}}-a^3\big(m_{\chi}^2+F(\phi)\Big)\chi_{\vec{k}}\chi_{-\vec{k}}\Big] .
  \end{equation}
   Using Eq.(\ref{3.3}) we can reach the following dynamical equation of each Fourier mode ($\chi_{\vec{k}})$ as,
   \begin{equation}\label{3.4}
      \ddot\chi_{\vec{k}}+3H\dot\chi_{\vec{k}}+\bigg(\frac{k^2}{a^2}+(m_{\chi}^2+F(\phi))\bigg)\chi_{\vec{k}}=0  .
  \end{equation}
  In the following dynamical Eq.(\ref{3.4}) there is a damping term, $3H\dot \chi_{\vec{k}}$ which is non-zero in expanding background. After rescaling the field $\chi_{\vec{k}}$ and defining $X_{\vec{k}}(t)= a^{\frac{3}{2}}\chi_{\vec{k}}(t)$, and we obtain,
   \begin{equation}\label{3.5}
      \ddot X_{\vec{k}}+\Omega_k^2(t)X_{\vec{k}}=0, ~~\mbox{with}~~~\Omega_k^2(t)=\bigg(\frac{k^2}{a^2}+m_{\chi}^2+F(\phi)-\frac{9}{4}H^2-\frac{3}{2}\dot H\bigg).
  \end{equation}
  Where $\Omega_k$ is identified as a time-dependent frequency. It is important to note that due to inflaton, the terms such as $F(\phi), H$, are oscillatory in nature. Influenced by these periodic driving force, each mode ($X_{\vec{k}})$ will be resonantly produced as particles. In the expanding background, the co-moving number density of such $\chi$ particles $n_k(t)$ with co-moving momentum mode $k$ can be expressed in a well-known formula \cite{Kofman:1994rk,Kofman:1997yn}
   \begin{equation}\label{3.7}
      n_k= \frac{\Omega_k}{2}\bigg(\frac{|\dot X_{\vec{k}}|^2}{\Omega_k^2}+|X_{\vec{k}}|^2\bigg)-\frac{1}{2}  .
  \end{equation} 
In order to implement the two-mode squeezed state formalism, we express the Lagrangian (\ref{3.3}) in terms of the rescaled field with two independent modes $(X_{\vec{k}}, X_{-\vec{k}})$,
 \begin{equation}\label{3.8}
      L_X=\frac{1}{2}\int \frac{d^3\vec{k}}{(2\pi)^3}\Big[\dot X_{\vec{k}}\dot X_{-\vec{k}}-\Big(\frac{k^2}{a^2}-\frac{9}{4}H^2-\frac{3}{2}\dot H+(m_{\chi}^2+F(\phi))\Big)X_{\vec{k}}X_{-\vec{k}}\Big], 
  \end{equation} 
and in terms of their canonically conjugate momenta, $ \Pi_{X_{-\vec{k}}}=\dot X_{\vec{k}}$ and $ \Pi_{X_{\vec{k}}}= \dot X_{-\vec{k}}$
the Hamiltonian will be  
   \begin{equation}\label{3.9}
       H_X=\frac{1}{2} \int\frac{d^3\vec{k}}{(2\pi)^3}\Big[\Pi_{X_{\vec{k}}}\Pi_{X_{-\vec{k}}}+\Big(\frac{k^2}{a^2}-\frac{9}{4}H^2-\frac{3}{2}\dot H+(m_{\chi}^2+F(\phi))\Big)X_{\vec{k}}X_{-\vec{k}}\Big] .
  \end{equation}
In the squeezed state formalism, the canonically conjugate phase space operators ($\hat{X}_{\vec{k}}(t), \hat{\Pi}_{X_{\vec{k}}}(t)$) are expressed terms of time-dependent creation and annihilation operators as
        \begin{align}\label{3.10}
          \hat{X}_{\vec{k}}&= \frac{1}{\sqrt{2\omega_k}}\big[\hat{a}_{\vec{k}}(t)+ \hat{a}_{-\vec{k}}^{\dagger}(t)\big]\nonumber\\
        \hat{\Pi}_{X_{\vec{k}}}&= -i \sqrt{\frac{\omega_k}{2}}\big[\hat{a}_{\vec{k}}(t)-\hat{a}_{-\vec{k}}^{\dagger}(t)\big] .
      \end{align}
Where time-independent frequency is defined as       $\omega_k=\sqrt{{k^2}/{a^2}+m_{\chi}^2}$. The time dependent creation and annihilation operators, $\hat{a}_{\vec{k}}(t)$ and $\hat{a}^{\dagger}_{-\vec{k}}(t)$ satisfy the standard commutation relation, $\big[\hat{a}_{\vec{p}},\hat{a}^{\dagger}_{-\vec{q}}\Big]= (2\pi)^3 \delta^3(\vec{p}+\vec{q})$. 
       Finally using Eq.(\ref{3.10}) in (\ref{3.9}) we are left with the desired form of the Hamiltonian of this system
       \begin{equation}\label{3.11}
          \hat{ H}_X=\frac{1}{2} \int\frac{d^3\vec{k}}{(2\pi)^3}\left[\left(\frac{\Omega_k^2(t)}{2\omega_k}+\frac{\omega_k}{2}\right)\left(\hat{a}_{\vec{k}}\hat{a}_{\vec{k}}^{\dagger}+\hat{a}_{-\vec{k}}^{\dagger}\hat{a}_{-\vec{k}}\right)+\left(\frac{\Omega_k^2(t)}{2\omega_k}-\frac{\omega_k}{2}\right)\left(\hat{a}_{\vec{k}}\hat{a}_{-\vec{k}}+\hat{a}_{-\vec{k}}^{\dagger}\hat{a}_{\vec{k}}^{\dagger}\right)\right] .
      \end{equation}
     In the above Hamiltonian, the time-dependent frequency $\Omega_k$ encodes the interaction between the daughter field and oscillatory inflaton. The first part of the Hamiltonian (\ref{3.11}) signifies the collection of free oscillators with equal and opposite momenta, and the second part is their interaction which creates or destroys them leading to well-known two-mode squeezing phenomena between them. Significantly, the classical time-dependent background is responsible for such phenomena. Once we obtained the required Hamiltonian, the Heisenberg equation is derived as,
      \begin{equation}\label{3.12}
      \frac{d}{dt}(\hat{a}_{\vec{k}})= i \big[\mathcal{A} \hat{a}_{\vec{k}}(t)+\mathcal{B} \hat{a}_{-\vec{k}}^{\dagger}(t)\big]
      \end{equation}
    where the new symbols are 
      \begin{equation}\label{3.13}
          \mathcal{A}=\Big(-\frac{\Omega_k^2(t)}{2\omega_k}-\frac{\omega_k}{2}\Big) \quad \text{and}\quad  \mathcal{B}=\Big(-\frac{\Omega_k^2(t)}{2\omega_k}+\frac{\omega_k}{2}\Big) .
      \end{equation}
      The Conventional notion tells us that time-dependent Hamiltonian of a system makes the concept of vacuum ambiguous. We shall justify this statement soon. In the present context, time dependence in the Hamiltonian (\ref{3.11}) comes through the interaction with oscillating inflaton background as well as expanding spacetime. However to solve the above Eq. (\ref{3.12}), we shall resort to the idea of Bogoliubov transformation,
       \begin{equation}\label{3.14}
          \hat{a}_{\vec{k}}(t)= \alpha_k(t)\hat{a}_{\vec{k}}(t_0)+\beta_k(t)\hat{a}_{-\vec{k}}^{\dagger}(t_0) .
      \end{equation}
      Where, $\alpha_k(t)$ and $\beta_k(t)$ are Bogoliubov coefficients satisfying the relation, $|\alpha_k(t)|^2-|\beta_k(t)|^2=1$ and $t_0$ is an initial time. The role of these Bogoliubov coefficients is to connect the late-time creation and annihilation operators with their counterparts at the initial time $t=t_0$. Using the Eq. (\ref{3.14}) in (\ref{3.12}) we obtain
      \begin{eqnarray}\label{3.15}
        && 
        \frac{d \alpha_k(t)}{dt}=i\mathcal{A}\alpha_k(t)+i\mathcal{B}\beta_k^{\star}(t) \\
     &&     \frac{d \beta_k(t)}{dt}=i\mathcal{A}\beta_k(t)+i\mathcal{B}\alpha_k^{\star}(t).
       \end{eqnarray}
 With a view to determining the expression of two Bogoliubov coefficients $\alpha_k$ and $\beta_k$, we shall introduce two operators following the reference \cite{Martin:2015qta, Haque:2020pmp}. The first one is two-mode unitary squeezing operator $\hat{S}_k(r_k,\varphi_k)$ defined as, $\hat{S}_k(r_k,\varphi_k)= e^{\hat{B}_k}$ with
       \begin{equation}\label{3.17}
           \hat{B}_k \equiv \frac{r_k(t)}{2}\Big( e^{-2i\varphi_k(t)}\hat{a}_{\vec{k}}(t_0)\hat{a}_{\vec{-k}}(t_0)- e^{2i\varphi_k(t)}\hat{a}_{\vec{-k}}^{\dagger}(t_0)\hat{a}_{\vec{k}}^{\dagger}(t_0)\Big).
       \end{equation}
  The unitarity of the squeezing operator implies the anti-hermitian nature of $\hat{B}_k^{\dagger}=-\hat{B}_k$ as can be seen from Eq.(\ref{3.17}). Two parameters $r_k$ and $\varphi_k$ in the Eq. (\ref{3.17}) are quantifying the amount of squeezing of the quantum state and its squeezing angle respectively. To have complete evolution of the system, the second operator is the rotation operator, $\hat{R}_k(\theta_k)$, defined as, $\hat{R}_k(\theta_k)= e^{\hat{D}_k}$ with
        \begin{equation}\label{3.18}
           \hat{D}_k \equiv -i\theta_k(t)\Big(\hat{a}_{\vec{k}}^{\dagger}(t_0)\hat{a}_{\vec{k}}(t_0)+\hat{a}_{\vec{-k}}^{\dagger}(t_0)\hat{a}_{\vec{-k}}(t_0)\Big),
       \end{equation}
where, $\hat{D}_k^{\dagger}=-\hat{D}_k$ as it was in Eq. (\ref{3.17}). Here rotation operator is characterized by a parameter $\theta_k$, called rotation angle. Combining these two operators $\hat{S}_k(r_k,\varphi_k)$ and $\hat{R}_k(\theta_k)$ one can construct a unitary time-evolution operator for every individual scalar mode as,
       \begin{equation}\label{3.19}
          \hat{ \mathcal{U}}_k= \hat{S}_k(r_k,\varphi_k)\hat{R}_k(\theta_k) .
       \end{equation}
One can easily check that the application of this unitary evolution operator on the creation operator generates its time evolution as
       \begin{equation}\label{3.20}
           \hat{a}_{\vec{k}}(t)= \hat{\mathcal{U}}_k^{\dagger}(t,t_0)\hat{a}_{\vec k}(t_0)\hat{\mathcal{U}}_k(t,t_0)
     \Rightarrow \hat{a}_{\vec{k}}(t)= \Big(\cosh r_k e^{-i\theta_k}\hat{a}_{\vec{k}}(t_0)- \sinh r_k e^{i(\theta_k+ 2\varphi_k)}\hat{a}_{\vec{-k}}^{\dagger}(t_0)\Big) .
       \end{equation}
A similar equation can be obtained for the annihilation operator by taking the complex conjugate. A comparison between Eq. (\ref{3.14}) and (\ref{3.20}) provides one of the forms of two  coefficients $\alpha_k(t)$ and $\beta_k(t)$ in terms of $r_k, \varphi_k$ and $\theta_k$ in this two-mode squeezed state formalism. The forms are as follows:
         \begin{eqnarray}\label{3.21}
    &&   \alpha_k(t)=e^{-i\theta_k}\cosh r_k \\
     &&           \beta_k(t)=-e^{(i\theta_k+2i\varphi_k)}\sinh r_k.
         \end{eqnarray}
  Such choice naturally satisfies the relation, $|\alpha_k(t)|^2-|\beta_k(t)|^2=1$. In light of this squeezed state formalism, we can alternatively define an expression of occupation number density $n_k$ which is equivalent to the expression (\ref{3.7}). It is worth mentioning that the purpose of defining such coefficients namely Bogoliubov coefficients is to connect the late-time raising and lowering operators with the operators of initial time. In Heisenberg picture, as operator evolution is studied rather than state evolution, a vacuum state corresponding to the operators at initial time will no longer be a vacuum corresponding to the creation and annihilation operators at late time. One can construct the late-time number operator($\hat{N}_{\vec{k}}(t)$) using (\ref{3.20}) and conjugate, $\hat{N}_{\vec{k}}(t)=\hat{a}_{\vec{k}}^{\dagger}(t)\hat{a}_{\vec{k}}(t)$ and the vacuum expectation value of this late-time number operator in the so-called Bunch-Davis vacuum defined at the initial time($t_0)$ can be expressed as,
         \begin{equation}\label{3.23}
             \langle \hat{N}_{\vec{k}}(t)\rangle= \bra{0_{\vec{k}}}\hat{a}_{\vec{k}}^{\dagger}(t)\hat{a}_{\vec{k}}(t)\ket{0_{\vec{k}}}=|\beta_k(t)|^2=\sinh^2 r_k
         \end{equation}
 The same expression can also be obtained for $-\vec{k}$ mode, $ \langle \hat{N}_{-\vec{k}}(t)\rangle= \sinh^2 r_k$.  Interesting to note that the evolution of squeezing parameter $r_k$ influences the time-evolution of occupation number density for various modes $\vec{k}$. A closer look at the equations (\ref{3.20}) signifies the fact that the mixing of raising and lowering operators in the evolution equation is solely responsible for a vacuum state in the remote past becoming an excited state in the distant future, hence justifying our previous statement regarding the ambiguity of the concept of vacuum for time-dependent Hamiltonian. 
\subsection{Two-mode squeezed states for bosonic particles} 
 We have already introduced key concepts of squeezed state formalism, and we proceed to identify the squeezed quantum state. It is a well-known fact that upon quantization, the states of a harmonic oscillator are coherent in nature and the states of a parametric oscillator(an oscillator having time-dependent frequency) are squeezed in nature. If one measures the uncertainty of two-phase space variables in a squeezed state corresponding to a particular system, then he will encounter the terms whereupon one grows exponentially and another one will die out exponentially \cite{Albrecht:1992kf,Hashimoto:2020xfr}. This can be clearly observed in an inverted harmonic oscillator(IO) state which is a beautiful example of a squeezed state. The Hamiltonian of IO having potential unbounded from below is given below:
          \begin{equation}\label{3.24}
        \hat{H}_{\rm IO}=\frac{\hat{p}^2}{2}-\frac{1}{2}k\hat{x}^2 .
         \end{equation} 
        By defining $\hat{x}$ and $\hat{p}$ in terms of raising and lowering operators of the non-inverted oscillator,
         \begin{equation}\label{3.25}
        \hat{x}=\frac{1}{\sqrt{2k}}\big(\hat{a}+\hat{a}^{\dagger}\big), \quad \hat{p}=-i\sqrt{\frac{k}{2}}\big(\hat{a}-\hat{a}^{\dagger}\big) ,
         \end{equation}
    the inverted oscillator Hamiltonian becomes
         \begin{equation}\label{3.26}
        \hat{H}_{\rm IO}=-\frac{k}{2}\big(\hat{a}^2+(\hat{a}^{\dagger})^2\big) .
         \end{equation}
         It can be easily shown that the squeezing of the quantum state for an inverted harmonic oscillator is due to the $(a^{\dagger})^2$ term. We notice that the second term of the Hamiltonian (\ref{3.11}) is the same as that of the inverted oscillator Hamiltonian in its field theoretic form. Hence, we can infer that the presence of background driving force due to inflaton, the Hamiltonian (\ref{3.11}) induces squeezing in the system.
         
          Introducing the fundamental concept for the well-known inverted harmonic oscillator, we shall now proceed towards the determination of the squeezed state by the application of the unitary evolution operator (\ref{3.19}) on a two-mode unsqueezed vacuum state. Borrowing the idea as outlined in \cite{Martin:2015qta}, we can factorize the full Hilbert space of the system $\boldsymbol{\varepsilon}$ into independent inner products of Hilbert spaces for two opposite momenta modes $\vec{k}$ and $-\vec{k}$,  $\boldsymbol{\varepsilon}=\prod_{k\in \mathbb{R}^{3+}}\boldsymbol{\varepsilon}_{\vec{k}}\otimes \boldsymbol{\varepsilon}_{-\vec{k}}$. Using this idea, we can write two-mode unsqueezed vacuum state as, $\ket{0_{\vec{k}},0_{-\vec{k}}} \equiv \ket{0_{\vec{k}}}\otimes \ket{0_{-\vec{k}}}$. Finally the application of two-mode unitary evolution operator (\ref{3.19}) on the unsqueezed vacuum leads us to the momentum-preserving two-mode squeezed state\footnote{One can explicitly check that the application of Rotation operator on the initial vacuum state keeps it invariant, $\hat{R}_{\vec{k}}(\theta_k)\ket{0_{\vec{k}},0_{-\vec{k}}}=  \ket{0_{\vec{k}},0_{-\vec{k}}}$, so vacuum state is rotationally invariant.}:
         \begin{equation}\label{3.27}
          \ket{\psi_{\rm sq}}_{\vec{k},-\vec{k}}= \hat{S}_k(r_k,\varphi_k)\hat{R}_k(\theta_k)\ket{0_{\vec{k}},0_{-\vec{k}}} =\frac{1}{\cosh r_k}\sum_{n=0}^{\infty}e^{2in\varphi_k}(-1)^n \tanh^n r_k\ket{n_{\vec{k}},n_{-\vec{k}}},
         \end{equation}
          where two-mode excited state is given by
         \begin{equation}\label{3.28}
      \ket{n_{\vec{k}},n_{-\vec{k}}}= \sum_{n=0}^{\infty}\frac{1}{n!}\big(\hat{a}_{\vec{k}}^{\dagger}\big)^n\big(\hat{a}_{-\vec{k}}^{\dagger}\big)^n\ket{0_{\vec{k}},0_{-\vec{k}}}
  \end{equation}
         This $\ket{\psi_{\rm sq}}_{\vec{k},-\vec{k}}$ is a two-mode squeezed state, well known in the context of quantum optics as an entangled state.
          \subsection{Dynamical equations of $r_k$, $\varphi_k$ and $\theta_k$ :}
          According to the equations (\ref{3.21}), we have the expression of two coefficients $\alpha_k(t)$ and $\beta_k(t)$ in terms of $r_k$, $\varphi_k$ and $\theta_k$. Plugging these expressions into the equations (\ref{3.15}), one obtains three dynamical equations of $r_k$, $\varphi_k$ and $\theta_k$. The utility of the solution of these dynamical equations lies in the study of not only the time evolution of the system but also the squeezed state (\ref{3.27}). Those equations are as follows:
         \begin{eqnarray}\label{3.29}
 &&    \dot r_k= -\mathcal{B}\sin 2\varphi_k \nonumber\\
 &&     \dot \varphi_k= \big(\mathcal{A}-\mathcal{B}\coth 2r_k \cos 2\varphi_k\big) \nonumber\\
&&    \dot \theta_k= -\big(\mathcal{A}-\mathcal{B}\tanh r_k\cos 2\varphi_k\big) .
 \end{eqnarray}
 Having analyzed the above three equations, we come to know that none of the dynamical equations of $r_k$, $\varphi_k$, and $\theta_k$ depend upon the rotation angle $\theta_k$. So while studying the dynamics, the system will not exhibit any such sensitive dependence on the rotation angle $\theta_k$. 
 \section{OTOC: A diagnostic of chaos}\label{sec4}
 OTOC, a nice acronym of {\it out-of-time-order correlator} is one of the diagnostics of specifying the presence of quantum chaos in a system. Unlike a classical system, studying and characterizing the chaos in quantum many-body systems is pretty challenging. For a classical system, if a small perturbation to the initial state of a system causes exponentially diverging phase-space trajectories, then such hypersensitivity to the initial state or initial condition of that system is a clear indication of the presence of chaos. This hypersensitivity can be quantitatively characterized by the Poisson bracket between position ($q$) and momentum ($p$) at unequal time:
 \begin{equation}\label{4.1}
     \{q(t),p(0)\}^2=\Big(\frac{\partial q(t)}{\partial q(0)}\Big)^2 \sim \sum_n c_n e^{2\lambda_n t}
 \end{equation} 
 Where, $\lambda_n$, the \textit{Lyapunov exponents} are known to quantify the measure the chaos. In a quantum mechanical system, an analogous quantity can be defined between two operators with the classical Poisson bracket replaced by a Commutation bracket. The quantum mechanical analog of (\ref{4.1}) is generically called OTOC $\big[\hat{q}(t),\hat{p}(0)\big]$, which behaves as a Poisson bracket in the semiclassical limit. Likewise, the exponential diverging trajectories in the classical case, the exponential amplification of this commutator is a potential signature of \textit{quantum chaos}. To quantify OTOC, the common practice is the use of a double unequal-time commutator
 \begin{equation}\label{4.2}
     \mathcal{C}(t)\equiv -\langle [\hat{q}(t),\hat{p}(0)]^2\rangle_{\beta}
 \end{equation} 
 Where, angle brackets $\langle...\rangle_{\beta}$ stands for thermal average and $\beta$ being the inverse temperature, $\beta={1}/(k_BT)$, where $k_B$ is the Boltzmann constant. Though we are taking commutation between $\hat{q}(t)$ and $\hat{p}(0)$, generically one can consider commutation between any pair of hermitian operators $\hat{W}$, $\hat{V}$,
 \begin{equation}\label{4.3}
    \mathcal{C}(t)\equiv -\langle [\hat{W}(t),\hat{V}(0)]^2\rangle_{\beta} . 
 \end{equation} 
Therefore, according to the above discussions, a quantum chaotic system is also expected to exhibit exponential growth in OTOC, $\mathcal{C}(t)\sim e^{2\lambda t}$, in a similar fashion to the classical one and we can define the analogous \textit{quantum Lyapunov exponent}($\lambda$) measuring chaos in the quantum system under consideration. In this work, we will calculate OTOC in the two-mode squeezed state parametrized described before for different classes of periodically driven systems in the context of preheating. For this we express the field and conjugate momentum (see Eqs. (\ref{3.10})) in terms of three squeezing parameters $(r_k, \varphi_k, \theta_k)$ as, 
  \begin{eqnarray}\label{4.4}
  &&   \hat{X}_{\vec{k}}(t)=\frac{1}{\sqrt{2\omega_k}}\big(\cosh r_ke^{-i\theta_k}-\sinh r_k e^{-i(\theta_k+2\varphi_k)}\big)\hat{a}_{\vec{k}}(t_0)+\frac{1}{\sqrt{2\omega_k}}\big(\cosh r_ke^{i\theta_k}-\sinh r_ke^{i(\theta_k+2\varphi_k)}\big)\hat{a}_{-\vec{k}}^{\dagger}(t_0) \nno\\     && \hat{\Pi}_{X_{\vec{k}}}(t)=-i\sqrt{\frac{\omega_k}{2}}\big(\cosh r_k.e^{-i\theta_k}+\sinh r_k.e^{-i(\theta_k+2\varphi_k)}\big)\hat{a}_{\vec{k}}(t_0)+i\sqrt{\frac{\omega_k}{2}}\big(\text{cosh}r_ke^{i\theta_k}+\sinh r_ke^{i(\theta_k+2\varphi_k)}\big)\hat{a}_{-\vec{k}}^{\dagger}.(t_0) \nno \\
    \end{eqnarray}
Using the above conjugate variables the commutation relation assumes the following simple form,
          \begin{equation}\label{4.5}
        \big[\hat{X}_{\vec{k}}(t), \hat{\Pi}_{X_{\vec{k}'}}(t_0)\big]= i.(2\pi)^3\delta^3(\vec{k}+\vec{k^{'}}).f_k(t,t_0),
           \end{equation}
           where  $f_k(t,t_0)$ is defined as
          \begin{equation}\label{4.6}
               f_k(t,t_0)= \frac{1}{2}\Big[\big(\cosh r_k.e^{-i\theta_k}-\sinh r_k.e^{-i(\theta_k+2\varphi_k)}\big)\big(\cosh r_0.e^{i\theta_0}+\sinh r_0.e^{i(\theta_0+2\varphi_0)}\big)+c.c\Big],
           \end{equation}
       with $(r_0, \theta_0, \varphi_0)$ being the values of the squeezing parameters  at initial time $t_0$. The advantage of using such a complex conjugate pair is its being a c-number operator (Eq.(\ref{4.5})). Using this property for every canonically conjugate pair of field variables in momentum space, we can straightforwardly compute the OTOC as 
          \begin{equation}\label{4.7}
        \mathcal{C}_{\vec{k}\vec{k}'}(t)=-\langle\big[\hat{X}_{\vec{k}}(t), \hat{\Pi}_{X_{\vec{k}'}}(t_0)\big]^2\rangle_{\beta} = (2\pi)^6 \mathcal{C}_{\vec{k}}(t)(\delta^3(\vec{k}+\vec{k}'))^2
           \end{equation}
          Clearly, the appearance of the delta function advocates the momentum conservation in the system which is reminiscent of a homogeneous FLRW background. Ignoring those delta function terms, paying attention only to the amplitude part, we have
          \begin{equation}\label{4.9}
            \mathcal{C}_{\vec{k}}(t) \sim f^2_{{k}}(t,t_0) .
          \end{equation} 
          In our subsequent studies, we will mainly focus on this amplitude part for various models. 
          We have already mentioned that not only the commutation between position and momentum operator but OTOC, in general, can be taken as a double-commutator between any two hermitian operators. Hence, all possible combinations between field and conjugate momentum can capture the behavior of OTOC. We, therefore, construct a $2\cross2$ matrix, $\mathcal{M}_{kk'}$,  which is the quantum counterpart of \textit{classical symplectic matrix} having unit determinant consisting of all such combinations of field momenta \cite{Gharibyan:2018fax,Haque:2020pmp}, as 
\begin{equation}\label{symplectic}
\mathcal{M}_{k k'}=  - \begin{pmatrix}
 \big[\hat{X}_{\vec{k}}(t),\hat{X}_{\vec{k}^{'}}(t_0)\big]^2 &  \big[\hat{X}_{\vec{k}}(t),\hat{\Pi}_{\vec{k}^{'}}(t_0)\big]^2 \\
 \big[\hat{\Pi}_{\vec{k}}(t),\hat{X}_{\vec{k}^{'}}(t_0)\big]^2 &  \big[\hat{\Pi}_{\vec{k}}(t),\hat{\Pi}_{\vec{k}^{'}}(t_0)\big]^2 
\end{pmatrix} =(2\pi)^6 (\delta^3(\vec{k}+\vec{k}'))^2 \begin{pmatrix}
    h_k^2(t,t_0) & f_k^2(t,t_0)\\
    g_k^2(t,t_0) & j_k^2(t,t_0)
\end{pmatrix} .
\end{equation}           
Where assuming the initial squeezing $r_0 \rightarrow 0$, all the components of the $c$-number matrix are calculated to be
\begin{eqnarray}\label{4.12}
&&     f_k(t,t_0)\approx \cosh r_k\cos (\theta_k-\theta_0)-\sinh r_k \cos (\theta_k-\theta_0+2\varphi_k)\nonumber\\
&&    g_k(t,t_0)\approx \cosh r_k \cos(\theta_k-\theta_0)+\sinh r_k \cos(\theta_k-\theta_0+2\varphi_k) \nonumber\\
&& h_k(t,t_0)\nonumber\approx \cosh r_k \sin(\theta_k-\theta_0)-\sinh r_k\sin(\theta_k-\theta_0+2\varphi_k)\nonumber \\
&&
     j_k(t,t_0)\approx \cosh r_k \sin(\theta_k-\theta_0)+\sinh r_k \sin(\theta_k-\theta_0+2\varphi_k) .
\end{eqnarray}
From the symplectic matrix, we shall then extract the behavior of OTOC from the dominant eigenvalue of this matrix, and extract the information of \enquote{Lyapunov exponent}($\lambda$) as discussed before. 

\section{Inflionary parameters and analytical solutions }\label{sec5}
To proceed further towards the OTOC calculation, first, we need to prepare the background which represents a coherently oscillating inflaton field, and the Hubble scale which plays the role of the periodic driving force for the daughter fields. A reminiscence of the Eq. (\ref{2.2}) enables us to determine the field dynamics as well as the Hubble scale numerically. As we are interested in three different equations of state, we study the background dynamics for three different $n=1,2,3$. To study background dynamics numerically, the specification of proper initial condition of field value just after inflation demands special importance. During inflation, the inflaton satisfies usual slow-roll conditions. The usual condition for the end of inflation is set by one of the slow roll parameters $\epsilon\propto \Big({V'}/{V}\Big)^2$ to be unity. In the context of $\alpha$-attractor E-type potential model, we can derive an expression of the field value for general $n$ at which inflation ends is
\begin{equation}\label{phiend}
      \phi_{\text{end}}=\sqrt{\frac{3\alpha}{2}}M_{\rm pl} \ln \Big(\frac{2n}{\sqrt{3\alpha}}+1\Big),
\end{equation}
 where \enquote{end} stands for the value at the end of inflation or at the beginning of reheating and this will be used as an initial condition in our full numerical solution of the inflaton in the post-inflationary phase. Using this field amplitude in (\ref{2.3}), we have the potential at the inflation end as
 
 \begin{equation}\label{Vend}
    V_{\text{end}}=\Lambda^4\bigg(\frac{2n}{2n+\sqrt{3\alpha}}\bigg)^{2n} 
 \end{equation}
 

\subsection{Analytics solutions during preheating}
In our work, we shall use the $\alpha$-attractor $E$-type potential potential model described earlier. Expanding the potential (\ref{2.3}) around its minima we get the approximated power-law type potential for general \enquote{$n$}, as
\begin{equation}\label{Vphin}
   V(\phi)\sim \Lambda^4\Big(\frac{2}{3\alpha}\Big)^{n}\Big(\frac{\phi}{M_{\rm pl}}\Big)^{2n} .
\end{equation}
In an expanding cosmological background, as the amplitude of the inflaton($\Phi(t)$) falls with time and generically assumes ${\phi}/{M_{\rm pl}}<<1$ for significant duration of the reheating phase. Hence, we can safely neglect the higher-order terms in the expansion in the potential. In the post-inflationary era, inflaton field oscillates around the minima of this approximated power-law potential (\ref{Vphin}) with a decaying amplitude $\Phi(t)$. Oscillatory behavior is strongly influenced by the nature of the potential near its minimum. At this stage, the generic solution of the first equation of (\ref{2.2}) can be expressed as, $\phi(t)=\Phi(t) \mathcal{P}(t) = \Phi(t) \sum_{\nu\ne 0} \mathcal{P}_\nu \cos(\nu \omega t)$, where $\Phi(t)$ is time-dependent inflaton amplitude in the dynamical background and $\mathcal{P}(t)$ is a quasi-periodic oscillatory function with the fundamental frequency for the periodic inflaton potential in Eq.(\ref{Vphin}) \cite{Garcia:2020wiy}, 
\begin{equation}\label{fundafre}
    \omega=\left(\sqrt{\frac{\partial^2 V}{\partial \phi^2}}\Big|_{\Phi(t)}\right)\sqrt{\frac{\pi n}{(2n-1)}}\frac{\Gamma\big(\frac{1}{2}+\frac{1}{2n}\big)}{\Gamma\big(\frac{1}{2n}\big)} .
\end{equation}
Note that the oscillation frequency is an explicit function of decaying inflaton amplitude. The energy-density, $\rho_{\phi}$ and pressure, $p_{\phi}$, of homogeneous scalar field inflaton can be written as
\begin{equation}\label{avgpphirhophi}
   \rho_{\phi}=\frac{1}{2}\dot{\phi}^2+V(\phi), \quad \text{and} \quad p_{\phi}=\frac{1}{2}\dot{\phi}^2-V(\phi) .
\end{equation}
Subject to the approximated power-law potential (\ref{Vphin}), ignoring background expansion and taking oscillation average over one complete oscillation, we have average energy density $ \langle\rho_{\phi}\rangle=V(\Phi)$, and pressure as $\langle p_{\phi}\rangle=w_\phi \langle\rho_{\phi}\rangle$. Where, the average post-inflationary EoS is expressed as, $w_{\phi}={(n-1)}/{(n+1)}$. In order to obtain decaying amplitude $\Phi(t)$ for general \enquote{$n$}, we utilise oscillation averaged energy-density and pressure in (\ref{avgpphirhophi}) together with (\ref{Vphin}) and the form of average EoS $w_{\phi}$ in the continuity equation, $\langle\dot{\rho}_{\phi}\rangle +3H(1+w_{\phi})\langle\rho_{\phi} \rangle \simeq 0$, this yields,
\begin{align}\label{phiampli}
    \dot{\Phi}=-\frac{3H}{n+1}\Phi 
   ~~ \Rightarrow ~~\Phi(a)=\phi_{\text{end}}\left(\frac{a}{a_{\text{end}}}\right)^{-\frac{3}{n+1}} .
\end{align}
Where $\phi_{\text{end}}$ and $a_{\text{end}}$ are inflaton amplitude and scale factor at the end of inflation respectively. Once the time-variation of inflaton amplitude is found for general \enquote{$n$}, using (\ref{phiampli}), we can express the leading order behavior of Hubble scale in terms of decaying amplitude and quasi-periodic oscillatory function for general EoS as follows.
\begin{equation}\label{hubbleforn}
    H(t)\simeq H_{\text{end}} \left(\frac{a}{a_{\text{end}}}\right)^{-\frac{3n}{n+1}}\Bigg(1+\frac{\mathcal{P}\sqrt{6(1-\mathcal{P}^{2n}})}{2(n+1)}\Big(\frac{\phi_{\text{end}}}{M_{\rm pl}}\Big) \left(\frac{a}{a_{\text{end}}}\right)^{-\frac{3}{n+1}}\Bigg) ,
\end{equation}
with $H_{\text{end}}=\sqrt{\frac{V_{\text{end}}}{2M^2_{\rm pl}}}$ being the inflationary energy scale at the end of inflation. We recover the well-known leading behavior of the Hubble parameter $\sim a^{-3n/(n+1)}$. As an example, for quadratic potential with $n=1$, we recover $H\sim a^{-3/2}$. Substituting (\ref{phiampli}) in $H^2\simeq {V(\Phi)}/{3M_{\rm pl}^2}$, we compute the leading order time variation of scale factor after inflation for general EoS.
\begin{equation}\label{scaleforn}
 a(t)=  a_{\text{end}} \left(\frac{m^{(n)}_{\rm eff} t}{m^{(n)}_{\rm eff} t_{\rm end}}\right)^{\frac{n+1}{3n}} =a_{\text{end}} \left(\frac{z_n}{z_n^{\rm end}}\right)^{\frac{n+1}{3n}}
\end{equation}
 We further define the dimensionless time variable $z_{n}={m^{(n)}_{\rm eff}}t/(2\pi)$ by multiplying effective inflaton mass parameter $m_{\rm eff}^{(n)}$ at the end of inflation.  
\begin{align}\label{meff}
  z_n^{\rm end}  = \frac{\sqrt{2n(2n-1)}(n+1)M_{\rm pl}}{2\pi\sqrt{3}n\phi_{\text{end}}} ~~;~~
  m_{\rm eff}^{(n)} =
  \sqrt{\frac{\partial^2 V}{\partial \phi^2}}\Bigg|_{\Phi =\phi_{\text{end}}} 
  \simeq \sqrt{2n(2n-1)}\Big(\frac{2}{3\alpha}\Big)^{\frac{1}{2}}\frac{\Lambda^{\frac{2}{n}}}{M_{\rm pl}} \big[V_{\text{end}}\big]^{\frac{n-1}{2n}}
\end{align}
 
In the present work, we will take the values $n=1$($w_{\phi}\simeq 0$), $n=2$($w_{\phi}\simeq 1/3$ ) and $n=3$($w_{\phi}\simeq {1}/{2}$). 

\subsubsection{ For $n=1$:}
\begin{figure}[t]
    \begin{center}
\includegraphics[scale=0.5]{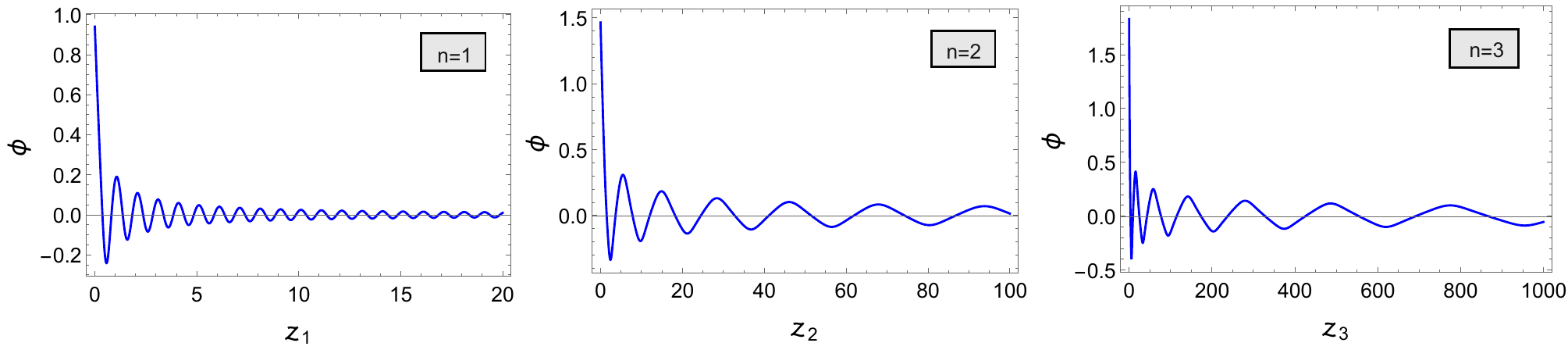}
\caption{\textit{Figure represents the post-inflationary evolution of inflaton background. Here field $\phi$ is taken in a unit of $M_{\rm pl}$ and we will be using these dimensionless time-variable $z_1, z_2, z_3$ throughout this work for $n=1,2,3$ respectively. } }
\label{phineq123}
\end{center}
\end{figure}

This is a model whereupon expanding the model potential about the minimum (i.e. at $\phi=0$) leads to quadratic power law potential ($ V \sim  \phi^2$).
From the expression (\ref{Vphin}) for $n=1$, we can obtain the inflaton mass $m^{(1)}_{\rm eff} = {2\Lambda^2}/{\sqrt{3\alpha}M_{\rm pl}}$ according to (\ref{meff})\big) which can be constrained by the CMB parameters($n_s,A_s$) through \enquote{$\Lambda$}. The form of that well-defined mass term is $m_{\phi}={2\Lambda^2}/{\sqrt{3\alpha}M_{\rm pl}}$  and subjected to the the central value of $n_s=0.9649$ and $A_s=(2.100\pm 0.030)\cross 10^{-9}$ we get $m^{(1)}_{\rm eff} \sim 1.23\cross 10^{-5}M_{\rm pl}$. It can be shown both analytically and numerically that at the end of inflation inflaton amplitude $\phi_{\text{end}}\sim M_{\rm pl}$, \and the condition $\sqrt{{2}/{3\alpha}}({\phi_{\text{end}}}/{M_{\rm pl}})<1$ gives $\alpha>0.5$. The latest observation suggests $\alpha \lesssim 10$. For our numerical purpose, we assume $\alpha=1$. The model corresponds to an average EoS $w_{\phi}\simeq 0$ and hence the background inflaton field behaves like non-relativistic matter.  
An approximate analytic solution can be recovered from Eq. (\ref{phiampli}) and
(\ref{hubbleforn}). 
\begin{equation}\label{anaphih1}
\phi(z_1)\approx\phi_{\text{end}}\Big(\frac{z_1^{\rm end}}{z_1}\Big)\cos (2\pi z_1), \quad H(z_1)\approx H_{\text{end}}\Big(\frac{z_1^{\rm end}}{z_1}\Big)\bigg(1+\frac{\sin(4\pi z_1)}{4\pi z_1}\bigg) .
\end{equation} 
For $n=1$, non-negligible contribution comes from only the first Fourier component $\mathcal{P}_1$, which is $|\mathcal{P}_1|=1/2$.   
From the above, we recover the leading order solution $a(t)\propto t^{2/3}$ implying a dust-dominated universe. The analytic solution can indeed be observed from the numerical solution presented in the left panel of Fig.\ref{phineq123}.  
 In our numerical study, we have got the value of inflaton field $\phi_{\text{end}}=0.94 M_{\rm pl}$ and Hubble scale $H_{\text{end}}=4\times 10^{-6}M_{\rm pl}$ at the end of inflation. 
\subsubsection{For $n=2$:}
For this case we have quartic power law potential $V\sim \overline{\lambda} \phi^4$ around the minimum, with $\overline{\lambda}=(16\Lambda^4)/(9\alpha^2M_{\rm pl}^4)
 \simeq 1.99\times 10^{-10}$ with $\alpha=1$.
Approximate analytic solution for $n=2$ obtained from Eq. (\ref{phiampli}) and
(\ref{hubbleforn}) as
 \begin{equation}\label{anaphi2}
\phi(z_2)\approx\phi_{\text{end}}\Big(\frac{z_2^{\rm end}}{z_2}\Big)^{\frac{1}{2}}\mathcal{P}(z_2), \quad H(z_2)\approx H_{\text{end}}\Big(\frac{z_2^{\rm end}}{z_2}\Big)\bigg(1+\frac{\mathcal{P}(z_2)\sqrt{6(1-\mathcal{P}^4(z_2))}}{6}\Big(\frac{\phi_{\text{end}}}{M_{\rm pl}}\Big)\Big(\frac{z_2^{\rm end}}{z_2}\Big)^{\frac{1}{2}}\bigg) .
 \end{equation}
 We numerically obtain $|\mathcal{P}_1|=0.47$ and $|\mathcal{P}_3|=0.02$. In terms of these Fourier components, $\mathcal{P}(z_2)$ takes the form $\mathcal{P}(z_2)\approx \Big(\mathcal{P}_1\cos \big(\frac{2\pi\omega}{m^{(2)}_{\rm eff}}z_2\big)+\mathcal{P}_3\cos \big(\frac{6\pi\omega}{m^{(2)}_{\rm eff}}z_2\big)\Big)$.
We indeed recover the leading order solution for the scale factor $a(t) \propto t^{1/2}$, which corresponds to average EoS $w_{\phi}\simeq 1/3$ implying radiation-like behavior. 
The numerical solution for inflaton is presented in the middle panel of Fig.\ref{phineq123}.
The value of inflaton field is obtained as  $\phi_{\text{end}}\sim 1.47 M_{\rm pl}$ and the Hubble scale $H_{\text{end}} \sim 3.6 \cross 10^{-6}M_{\rm pl}$ at the end of inflation. 
 \subsubsection{For $n=3$:}
For this case we have sextic type potential $V\sim \phi^6$, and approximate analytic solution derived from Eq. (\ref{phiampli}) and
(\ref{hubbleforn}), are
 \begin{equation}\label{anaphi3}
    \phi(z_3)\approx\phi_{\text{end}}\Big(\frac{z_3^{\rm end}}{z_3}\Big)^{\frac{1}{3}}\mathcal{P}(z_3), \quad H(z_3)\approx H_{\text{end}}\Big(\frac{z_3^{\rm end}}{z_3}\Big)\bigg(1+\frac{\mathcal{P}(z_3)\sqrt{6(1-\mathcal{P}^6(z_3))}}{8}\Big(\frac{\phi_{\text{end}}}{M_{\rm pl}}\Big)\Big(\frac{z_3^{\rm end}}{z_3}\Big)^{\frac{1}{3}}\bigg)
 \end{equation}
We numerically obtain $|\mathcal{P}_1|=0.46$ and $|\mathcal{P}_3|=0.03$. In terms of these Fourier components, $\mathcal{P}(z_3)$ takes the form $\mathcal{P}(z_3)\approx \Big(\mathcal{P}_1\cos \big(\frac{2\pi\omega}{m^{(3)}_{\rm eff}}z_3\big)+\mathcal{P}_3\cos \big(\frac{6\pi\omega}{m^{(3)}_{\rm eff}}z_3\big)\Big)$.  
We again recover the leading order scale factor $a(t) \propto t^{4/9}$, implying a stiff background with $w_{\phi}\simeq 1/2$. The numerical solution for the inflaton is presented in the third panel of  Fig.\ref{phineq123}, and we obtained $\phi_{\text{end}} \sim 1.8 M_{\rm pl}$ using (\ref{phiend}) and Hubble scale $H_{\text{end}}\sim 3.5 \times 10^{-6}M_{\rm pl}$ at the end of inflation.

Our study will be based on two different interactions, four-leg and three-leg type, and for each of these two, we will use three different background equations of state corresponding to $n=1,2,3$ in the potential (\ref{2.3}). 

\section{Introducing Interaction}\label{sec6}
Now we are on the verge of applying the above formalism in the context of the early reheating era, well-known as preheating. During this stage, the inflaton field sets the background as a periodic driving force as just discussed above, and acts as a source for the massless daughter field. Our main objective of this paper would be to understand in detail the dynamical structure of this massless production process in order to gain some insight into the quantum chaos under periodic driving force in the cosmological setting.

 In the present work, we are going to deal 
 two types of interactions: i) Three-leg interaction and ii) Four-leg interaction with the massless daughter field  say $\chi$,
\begin{equation}\label{threelegi}
    \mathcal{L}_{int}=\frac{1}{2}\sigma \phi \chi^2 ~~;~~
    \mathcal{L}_{int}=\frac{1}{2} g^2 \phi^2 \chi^2 
\end{equation}
where $(\sigma,g)$ are the coupling constant. In the present study, the generic coupling function of inflaton, $F(\phi)$, will have the forms $F(\phi)=\sigma\phi,g^2\phi^2$ corresponding to three-leg and four-leg type interaction respectively. In the usual framework of perturbative field theory such interaction leads to the decay of inflaton $\phi \rightarrow \chi\chi, \phi \phi \rightarrow \chi\chi $. We consider coupling parameter regime where the $\chi$-particle production will be non-perturbative \cite{Kofman:1997yn,Lozanov:2019jxc,Amin:2014eta}. However, such non-perturbative production stops as time progresses. In our following discussion, we discuss the important condition under which the system will be in the non-perturbative resonance regime with the aforementioned inflaton interaction. 
\subsection{Dynamical resonance condition: stability-instability chart}

\begin{figure}[t]
    \begin{center}
\includegraphics[scale=0.32]{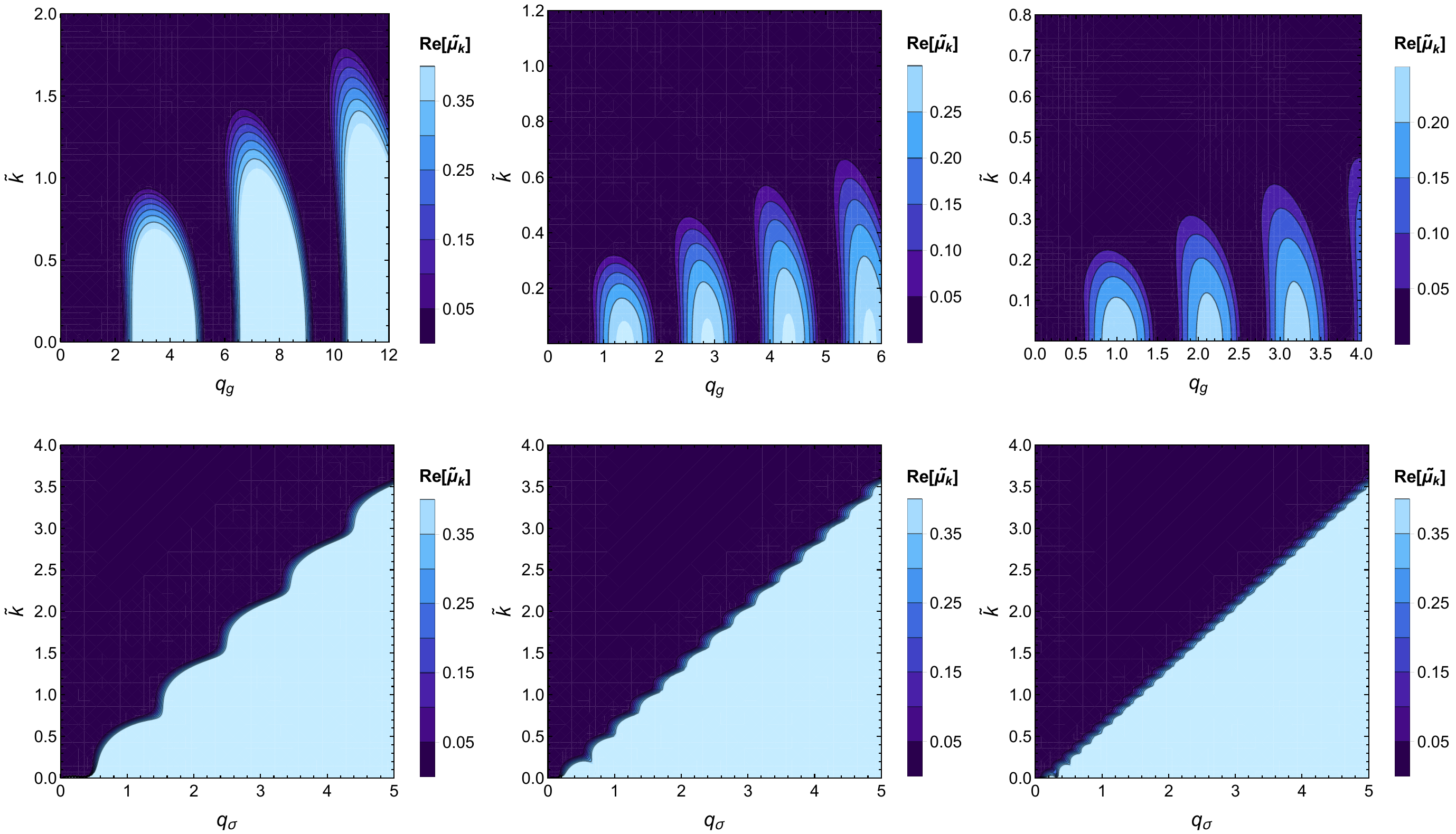}
\caption{\textit{\textbf{Upper Panel:} Figure represents the flat space parameter space for the model $n=1$, $n=2$ and $n=3$ respectively from the left corresponding to  $g^2\phi^2\chi^2$ interaction. \textbf{Lower Panel :} Figure represents the flat space parameter space for the model $n=1$, $n=2$ and $n=3$ respectively from the left corresponding to  $\sigma\phi\chi^2$ interaction.    }}
\label{parameterspace}
\end{center}
\end{figure}

While constructing parameter space to study resonance phenomenon during the early reheating phase, one usually considers static background ($a(t)=1$) neglecting the red shifting of momentum mode because of background expansion. Depending upon a specific background we are to construct this parameter space.
The generic mode equation assumes the following form,
\be
\frac{d^2 X_k}{dz_n^2} + \frac {4\pi^2}{m_{\rm eff}^{(n)2}}\left(\frac{k^2}{a^2 }-\frac{9}{4}H^2-\frac{3}{2}\dot H+ F(\phi) \right) X_k=0
\ee
As stated earlier the oscillating inflaton field is expressed as $\phi(z_n) = \Phi \mathcal{P}(z_n)$ with amplitude, $\Phi$. This is the well-known generalized Hill equation \cite{magnus1966hill}. The solution of such an equation is strongly dependent upon two parameters $(k,q)$. Where the emergent q-parameter, identified with the term associated with the purely oscillatory part of the equation is called resonance strength parameter, and in a static background those are,
\begin{equation}\label{resonq}
  q_g = \frac{g\phi_{\rm end}}{m_{\rm eff}^{(n)}}~ ~~{\rm for}~~\phi^2 \chi^2~~~~;~~~~
q_{\sigma} = \sqrt{\frac{\sigma \phi_{\rm end}}{m_{\rm eff}^{(n)2}}} ~~{\rm for}~~\phi \chi^2 
\end{equation}
Employing {\em Floquet theorem} \cite{magnus1966hill,mclachlan1947theory}, solution of such  equation shows exponential instability $X_k\propto exp(\tilde \mu_k z_n)$ over time for a specific range of parameter values ($k,g$) forming a band for which $Re\big(\tilde\mu_k \big)>0$, where $\tilde \mu_k$ is the rescaled Floquet exponent, $\tilde \mu_k={2\pi \mu_k}/{m^{(n)}_{\rm eff}}$. Generating a parameter space we can have those sets of different values of the two parameters($k$, $g$) from the unstable region for which the field solution will exhibit resonant instability. Consideration of a static Minkowskian background shows a clear distinction between stability and instability region in the parameter space (see Fig.\ref{parameterspace}). Any parameter set chosen from the instability region will lead to an unstable solution interpreted as resonant particle production, and vice versa. In reality, though, the background is dynamic, and hence parameters particularly the $k$ mode instead of staying in a specific stability/instability band may transit from one band to another with the expansion. 
Thus in the case of expanding background, the distinction between stability and instability region disappears which is a familiar characteristic of \textit{Stochastic resonance}. Because of the redshifting of each $k$ mode as well as small coupling strength $g$, if we choose any mode from a narrow resonance regime in parameter space, we will not be able to see the resonance phenomenon as it turns out to be almost instantaneous. Therefore, to observe the exponentially growing solution in time, we need to choose a parameter set from a broad resonance regime with sufficiently high coupling strength $g$. Employing the usual Floquet analysis for the periodically oscillating background, we have constructed these parameter spaces as shown in Fig.\ref{parameterspace}
for three different models corresponding to background $n=1,2,3$ for two different interactions. The structural difference in parameter space for two different interactions is very evident from the given figure. \textit{Tachyonic instability}\cite{Dufaux:2006ee,Felder:2001kt}, due to switching sign of $\mathcal{P}(z)$ for tri-linear coupling, causes a very efficient and strong resonance, and because of that, in the lower panel of Fig.\ref{parameterspace}, we notice a very broad instability region in the parameter space for the three given model in comparison with quartic or four-leg interaction.  

In the discussion so far we have talked about distinctive properties of flat-space resonance and resonance in the non-static background. But one thing that is not specified yet, is the precise condition of resonance for any general $n$. In the literature, conditions of resonance are usually specified by \cite{Kofman:1997yn,Greene:1997fu,Greene:2000ew}, considering $q>1$, and that leads to broad resonance. Otherwise, $q<1$ gives production in a very narrow resonance regime. Strictly speaking, such a condition defined in flat space is not appropriate, particularly when the driving force namely the inflaton dilutes very fast, $\Phi\propto \phi_{\text{end}} ({a}/{a_{\text{end}}})^{-{3}/{n+1}}$ depending on $n$ values. In the background of such a fast-falling driving force, the resonance parameter $q$ also reduces very rapidly in time as follows
\begin{align}\label{decayresonq2}
   q_{g}(z_n)= \frac{g\phi_{\text{end}}}{m_{\rm eff}^{(n)}} \left(\frac{z_n}{z_n^{\rm end}}\right)^{-\frac{1}{n}}
   ~~;~~
   q_{\sigma}(z_n)= \sqrt{\frac{\sigma \phi_{\text{end}}}{m_{\rm eff}^{(n)2}}} \left(\frac{z_n}{z_n^{\rm end}}\right)^{-\frac{1}{2n}}
\end{align}
For those cases, the naive Minkowski resonance condition indeed overestimates the lower limit of the inflaton coupling parameters above which broad-resonance could occur \footnote{We have numerically verified this statement}. For the present case with a rapidly falling $q$-parameter, therefore, {\it we propose a dynamical condition of resonance as follows:}

Resonant particle production is intimately tied with the violation of the adiabaticity condition, and that is expected to occur while inflaton crosses zero during its course of oscillation. To achieve significant resonant production within a certain period one needs to satisfy two important conditions. The first one would be to have the driving force, the oscillating inflaton, executing a few oscillations within the period of interest. The second condition is that within that period the resonance $q$-parameter should remain greater than unity. Combining the preceding two conditions, we state that \textit{\it{for broad resonance to take place while the resonance parameter $q$ time-evolving from its higher initial value to unity, it must complete at least one oscillation.}}  Bearing this dynamical condition of broad resonance in mind, we derive the lower bound of the inflaton couplings for general background EoS.

To compute the number of oscillations required for $q$-parameter to change from its large initial value at the end of inflation to unity, we measure the dimensionless time-period of the oscillating inflaton as $z_n^{(\omega)} = m_{eff}^{(n)}/\omega$, where $\omega$ is calculated at $\Phi = M_{pl}$. With respect to this, the number of oscillations say $N_{\rm osc}$ becomes,
\begin{align}\label{noosci}
     N_{\text{osc}}= \frac{z_n -z_n^{\rm end}}{z_n^{(\omega)}} = 
   \begin{cases}
  & \frac{z_n^{\rm end}}{z_n^{(\omega)}} \left(\left(\frac{g\phi_{\text{end}}}{m_{\rm eff}^{(n)}}\right)^n-1\right)\\
   & \frac{z_n^{\rm end}}{z_n^{(\omega)}} \left(\left(\frac{\sqrt{\sigma\phi_{\text{end}}}}{m_{\rm eff}^{(n)}}\right)^{2n}-1\right)
   \end{cases} .
\end{align}
This dimensionless time-period $z_n^{(\omega)}$ matches very well with the numerically calculated one. Hence, to achieve efficient resonant production in a broad resonance regime, the minimum criterion that must be fulfilled is $N_{\rm{osc}}>1$. Having this criterion, from (\ref{noosci}) we get the lower bound of the coupling parameter for two interactions as \cite{Chakraborty:2023ocr}
\begin{eqnarray}\label{couplingbound}
g &>& \frac{m^{(n)}_{\rm eff}}{\phi_{\text{end}} }\bigg[1+ \frac{z_n^{(\omega)}}{z_n^{\rm end}}\bigg]^{\frac{1}{n}}\nonumber\\
 \sigma &>& \frac{m^{(n)2}_{\rm eff}}{\phi_{\text{end}} }\bigg[1+\frac{z_n^{(\omega)}}{z_n^{\rm end}}\bigg]^{\frac{1}{n}} .
\end{eqnarray}
It is very interesting to note that the lower limit of the coupling strength $g$ to achieve efficient resonance also depends upon the time period of background oscillation besides its dependence on $\phi_{\text{end}}$ and $m^{(n)}_{\rm eff}$. Therefore this limiting value varies from one EoS to another depending upon $\phi_{\text{end}}, m^{(n)}_{\rm eff}$ and time-period $z_n^{(\omega)}$.

The above condition sets the minimum possible coupling parameter that one requires at the beginning of the reheating to have efficient broad parametric resonance. However, due to expanding background, any particular resonant mode $k$ also becomes non-resonant around its characteristic time scale. For example, in the present context, minimum dimensionless coupling strengths which are necessary to initiate an efficient resonance process are found to be $g=\sigma/m^{(n)}_{\rm eff}=6\times 10^{-5},4\times 10^{-5},3.77\times 10^{-5}$ for $n=1,2,3$ respectively. It is important to mention that these lower bounds of coupling parameters are obtained for parametric resonance cases. But for three-leg type interaction, efficient production happens for coupling strength lower than the given bound because of Tachyonic growth present in the system.\\ In the next section, while studying OTOC, we shall choose coupling parameters obeying the bound (\ref{couplingbound}).    


\section{Squeezing and OTOC: for four-leg interaction ($g^2\phi^2\chi^2$):}\label{sec8}
In this section, we will concentrate on the study of chaos by explicitly calculating OTOC for $n=1,2,3$ for four-leg or quartic interaction $(g^2\phi^2\chi^2)$. The daughter quantum field $\chi$ which experiences periodic driving force due to a coherently oscillating inflaton field, will be observed to undergo resonant amplification. Such resonant amplification will be shown to be chaotic in nature. As discussed before, we compute OTOC on the time-evolving squeezed quantum state and show that such quantity will exponentially grow with time. The inverse time scale of such exponential growth will be identified as the Lyanupov exponent. We shall discuss the chaotic scenario and its multifarious characteristics with minute details for three different background dynamics as constructed in the last section separately. 
To this end, we would like to point out that in our present analysis, we will confine ourselves strictly to the regime where back-reaction is negligible. 
\subsection{Initial condition specification:}
To perform a detailed numerical analysis the key dynamical equations are given in (\ref{3.29}). 
For the appropriate initial conditions, we first write down the field mode in terms of three squeezing parameters. Replacing Eq. (\ref{3.10}) in the mode expansion form of rescaled field $\hat{X}(t,\vec{x})$ like (\ref{3.2}) and also using (\ref{3.20}), we get the following form
\begin{eqnarray}\label{fluctuf}
  \hat{X}(t,\vec{x}) &=& \int\frac{d^3\vec{k}}{(2\pi)^3}.\frac{1}{\sqrt{2\omega_k}}\bigg[e^{-i\theta_k}\big(\cosh r_k-e^{-2i\varphi_k}\sinh r_k\big)\hat{a}_{\vec{k}}(t_0).e^{i\vec{k}.\vec{x}}+e^{i\theta_k}\big(\cosh r_k-e^{2i\varphi_k}\sinh r_k\big)\hat{a}_{-\vec{k}}^{\dagger}(t_0).e^{-i\vec{k}.\vec{x}}\bigg] \nno\\
  &=& \int\frac{d^3\vec{k}}{(2\pi)^3}.\big[X_k(t)\hat{a}_{\vec{k}}(t_0).e^{i\vec{k}.\vec{x}}+X_k^{\star}(t)\hat{a}_{\vec{k}}^{\dagger}(t_0).e^{-i\vec{k}.\vec{x}} .
\end{eqnarray} 
The general field mode function can further be expressed in terms of squeezing parameters $r_k,\varphi_k, \theta_k$ as,
\begin{equation}\label{fluctusqz} X_k(t)=\frac{e^{-i\theta_k}}{\sqrt{2\omega_k}}\big[\cosh r_k-e^{-2i\varphi_k}\sinh r_k\big]
\end{equation}
The proper specification of the initial condition can be associated with the well-known Bunch-Davies for each individual mode fluctuation. We can assume the solution of Eq. (\ref{3.5}) by WKB ansatz of the form
\begin{equation}\label{wkb}
    X_k(t)=\frac{1}{\sqrt{2\Omega_k}}\bigg[\alpha_k(t)e^{-i\int^{t}_{0}\Omega_k(t^{'})dt^{'}}+\beta_k(t)e^{i\int^{t}_{0}\Omega_k(t^{'})dt^{'}}\bigg] .
\end{equation}
Now, we can choose the form of $\alpha_k(t)$ and $\beta_k(t)$ as
\begin{equation}\label{alphabeta}
   \dot \alpha_k=\frac{\dot\Omega_k}{2\Omega_k}e^{-2i\int^{t}_{0}\Omega(t^{'})dt^{'}}\beta_k(t),  \dot \beta_k=\frac{\dot\Omega_k}{2\Omega_k}e^{2i\int^{t}_{0}\Omega(t^{'})dt^{'}}\alpha_k(t)
\end{equation}
with the condition
\begin{equation}\label{alphabetaini}
    \alpha_k(t=0)=1,\quad  \beta_k(t=0)=0
\end{equation} 
to achieve  Bunch-Davies vacuum in the field mode $X_k$ 
\begin{equation}\label{bunchdaviesf}
    X_k(t\rightarrow 0)\simeq\frac{1}{\sqrt{2\Omega_k}}e^{-i\Omega_kt}=\frac{1}{\sqrt{2\Omega_k}}(\cos \Omega_kt-i\sin \Omega_kt)
\end{equation} 
and it's derivative is
\begin{equation}\label{bunchdaviesd}
  \dot X_k(t\rightarrow 0)\simeq-i\sqrt{\frac{\Omega_k}{2}}e^{-i\Omega_kt}=-i\sqrt{\frac{\Omega_k}{2}}(\cos \Omega_kt-i\sin \Omega_kt)  
\end{equation}
Comparing Eq. (\ref{alphabetaini}) with (\ref{3.21}) at the initial time $t=0$, we get the initial conditions for $r_k(t=0)=0$, and $\theta_k(t=0)= 2 l \pi$, where $l$ is any positive integer including zero. We have the freedom to take any value of $\varphi_k(t=0)$ since all the physical quantities are independent of squeezing angle $\varphi_k$. Such choices essentially make the initial condition for the initial quantum state to be an unsqueezed one, and that is equivalent to the choice of Bunch-Davies vacuum. Once initial conditions are properly specified we are now in a position to proceed with our numerical studies.
\subsection{Computing OTOC for $n=1$}
\begin{figure}[t]
 \begin{center}
 \includegraphics[scale=0.28]{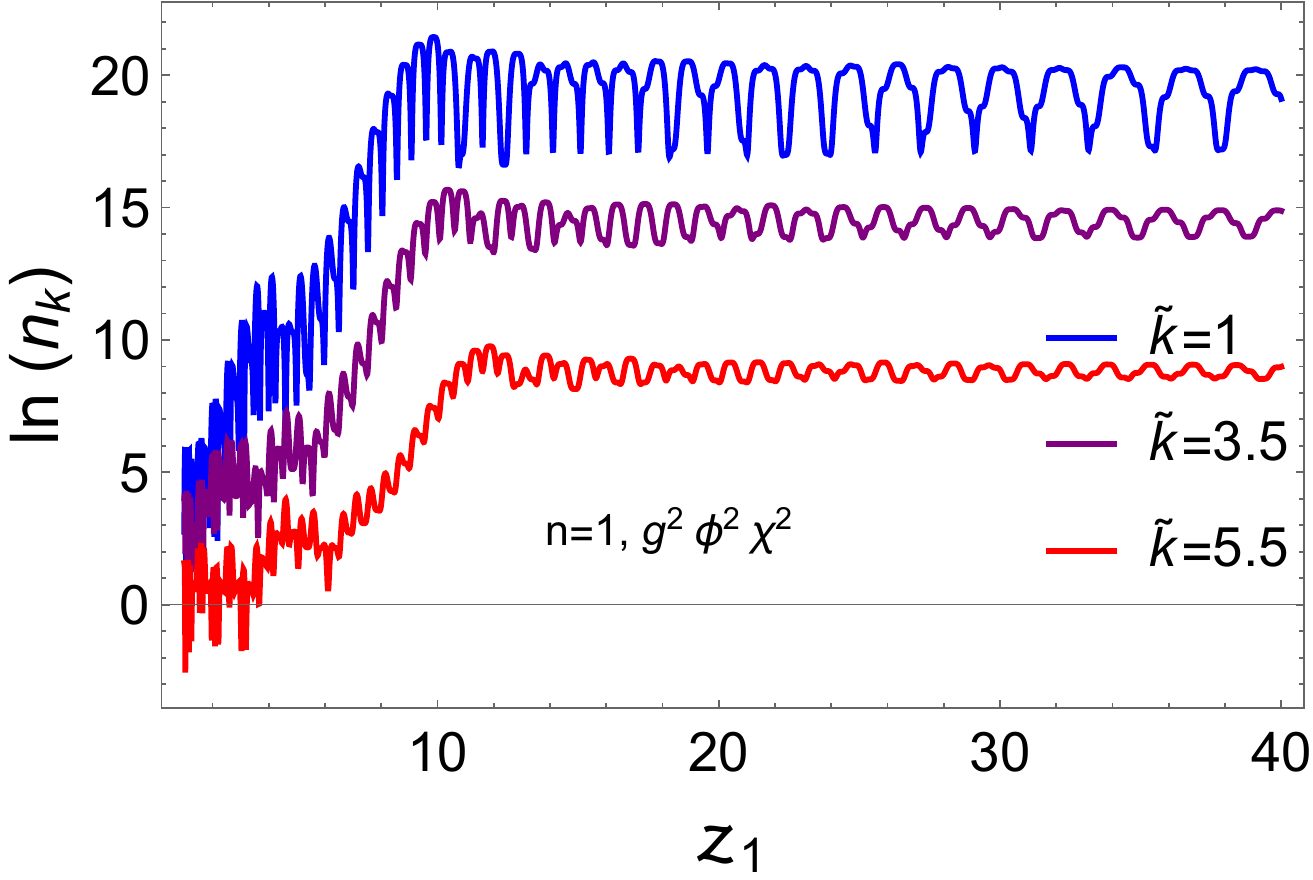}
\includegraphics[scale=0.42]{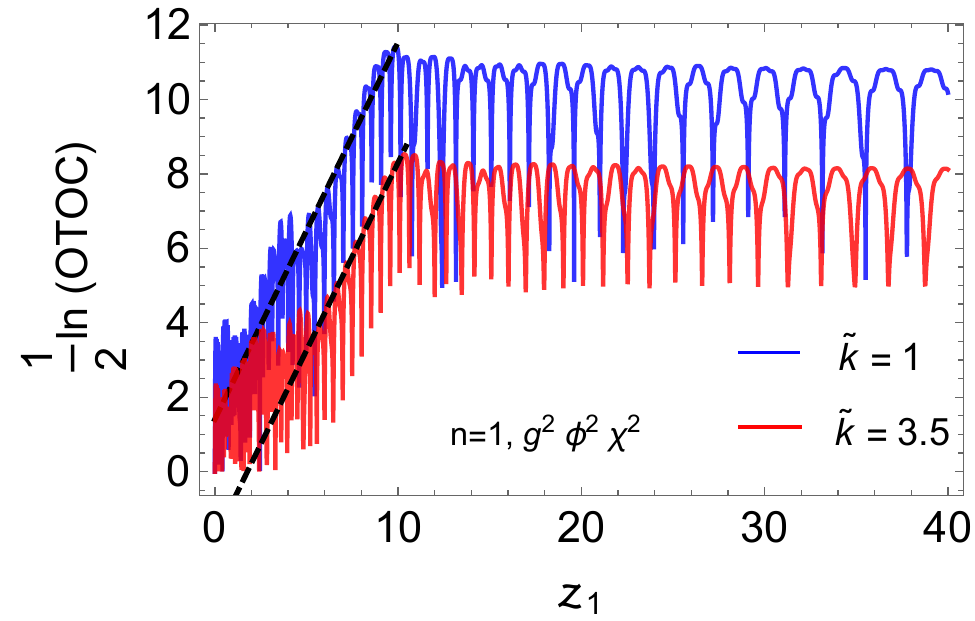} 
\caption{\textbf{Left Panel:}\textit{Figure represents the growth of occupation number density $n_k$ for three efficient $k$ modes with $g=5\times 10^{-4}$.} \textbf{Right Panel:} \textit{Figure represents the time variation of logarithm of OTOC amplitude for two different $k$ modes corresponding to significant production with $g=5\times10^{-4}$. Two black dashed lines for two modes indicate the average straight line-like growth of} ln(OTOC) \textit{and from the slope of those lines we estimate the average growth indices of chaos of the system.}}
\label{lnOTOCbetaneq1qua}
\end{center}
\end{figure}

We shall now present our numerical results for $r_k,\varphi_k, \theta_k$  solving three first-order coupled differential equations (\ref{3.29}). We will be interested in those parameter ($k,g$) that belong to the instability region of the stability-instability chart (see Fig.\ref{parameterspace}), and hence the resonant solution exists. It is a well-known fact that long-wavelength modes (IR modes) are efficiently amplified during the preheating era. For pictorial representation and comparison, we have chosen three different long wavelength modes and inflaton daughter field coupling constant $g=5\times 10^{-4}$.   
\begin{figure}[t]
     \begin{center}
\includegraphics[scale=0.55]{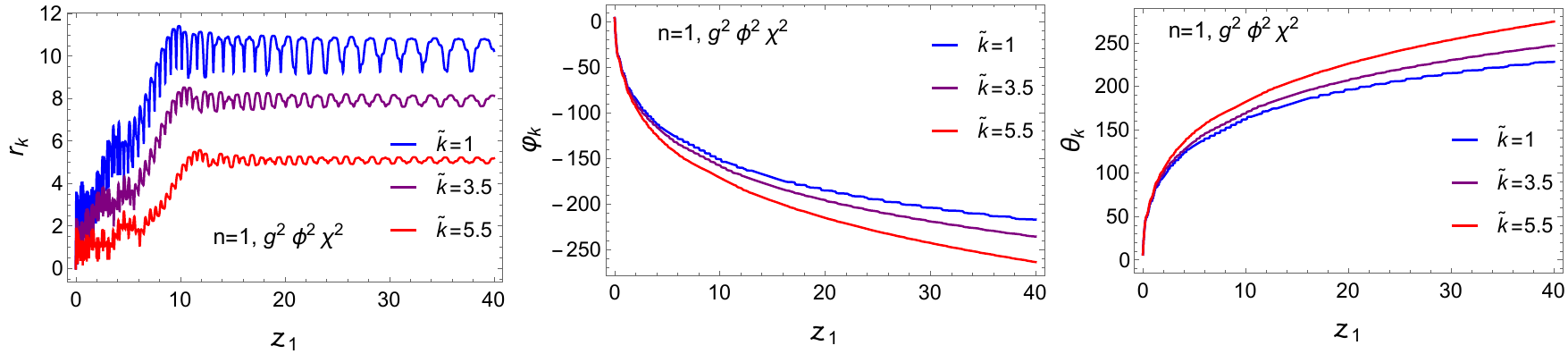}
\caption{\textit{Figure represents the time evolution of squeezing parameter$(r_k)$, squeezing angle$(\varphi_k)$ and rotation angle$(\theta_k)$ for three different momentum modes$(k)$ and the coupling strength is chosen to be $g=5\times 10^{-4}$. Here Squeezing angle($\varphi_k$) and rotation angle ($\theta_k$) are given in units of radian. Here we have taken three different modes for which squeezing parameters grow efficiently giving a considerable production of particles.   }}
\label{sqzparaneq1qua}
\end{center}
\end{figure}
With those parameter choices, we solve for the squeezing parameters shown in Fig.\ref{sqzparaneq1qua}. During the initial phase of the evolution, the squeezing parameter, $r_k$ grows linearly in time $z_1$ and subsequently saturates. As expected, the slope of the growth increases with decreasing momentum. Long wavelength mode takes longer time to saturate, that depends on the decreasing amplitude of the background driving force due to Hubble expansion. All these features can indeed be seen in Fig.\ref{sqzparaneq1qua}. Subject to the variation of $r_k$ as shown in Fig.\ref{sqzparaneq1qua} and using Eq. (\ref{3.23}), we obtain occupation number density, $n_k$ (See Fig.\ref{lnOTOCbetaneq1qua}), grows exponentially in time for all the three modes under consideration. However, generic dynamical features of squeezing ($r_k$) and number density $\text{ln}(n_k)$ are that they initially grow and reaches a peak value and finally the system relaxes to saturation with small periodic fluctuations. Furthermore, as expected the long wavelength modes are easily excited than those of the higher momentum mode. These features nicely resemble the thermalization process of any thermodynamic system under perturbation. For every individual mode, the thermalization associated with the saturated region can be said to be achieved at their respective time scale. Long wavelength modes thermalize faster in time than short wavelength ones. From Fig.\ref{sqzparaneq1qua} we obtain different values of saturation time scale for three different modes having different wavelengths. For $k=m^{(1)}_{\rm eff}$, we have $z_1\sim 9.9$, for $k=3.5m^{(1)}_{\rm eff}$, we have $z_1\sim 10.7$ and for $k=5.5m^{(1)}_{\rm eff}$, we have $z_1\sim 12$.

For the present system, the reason behind such variation of saturation time scale is hidden in the dynamics of the background periodic driving force due to the coherently oscillating inflaton field. The dimensionless resonance parameter which controls the dynamics is $q_{g}^2= (g/m^{(1)}_{\rm eff})^2 \Phi(t)^2 $.  Where, $\Phi(t)$ is decaying as $\sim a^{-3/2}$ (see Eq.(\ref{anaphih1})). The condition of stochastic resonance in expanding background must satisfy \cite{Kofman:1997yn}, $q_{g}^4m^{(1)}_{\rm eff}\gtrsim H$, we have
\begin{equation}\label{stochasticcon}
    g\Phi(t)\gtrsim m^{(1)}_{\rm eff} \bigg(\frac{H_{\rm end}}{m^{(1)}_{\rm eff}}\bigg)^{\frac{1}{4}} \left(\frac {z_1}{z_{1}^{\rm end}}\right)^{-\frac{1}{4}}.
\end{equation}
Considering CMB observed value of scalar spectral index $n_s = 0.9649 \pm 0.0042$, we have $H_{\rm end}\sim 10^{-6}M_{\rm pl}$. Post-inflationary dynamics leads ${H}/{m^{(1)}_{\rm eff}}=(H_{\rm end}/m^{(1)}_{\rm eff}) (z_{1}^{\rm end}/z_1) $ very small throughout the entire phase with $m^{(1)}_{\rm eff}\sim 10^{-5} M_{\rm pl}$. 
On the other hand, the amplitude of $\Phi$ also decreases, $\Phi\propto z_1^{-1}$ with the background expansion. Therefore, we have competition between the aforesaid two effects. As the dependence of the above condition (\ref{stochasticcon}) on $H$ is weak ($H^{1/4}$), It has been observed that 
after a certain number of inflaton oscillations the above resonance condition (\ref{stochasticcon}) fails to satisfy and eventually resonance process ceases. This causes the saturation in all the physical quantities namely occupation density $(n_k)$, squeezing amplitude $(r_k)$, etc as shown in Fig.\ref{sqzparaneq1qua} and Fig \ref{lnOTOCbetaneq1qua}.

So far we discussed in detail the dynamics of quantum states of the produced particles in terms of their squeezing parameters and phenomena of chaotic resonance in terms of the field modes. In the following, we will study in detail how such non-trivial behavior of resonance is encoded into OTOC considered as an interesting diagnostic tool of chaos. We first numerically construct the symplectic matrix in terms of commutators of field mode and its conjugate momentum. From the dominant eigenvalue of the calculated symplectic matrix (\ref{symplectic}) we identify the OTOC dynamics which is parametrized by the $Lyapunov$ $exponent$($\lambda_k$). 

\begin{figure}[t]
    \begin{center}
\includegraphics[scale=0.550]{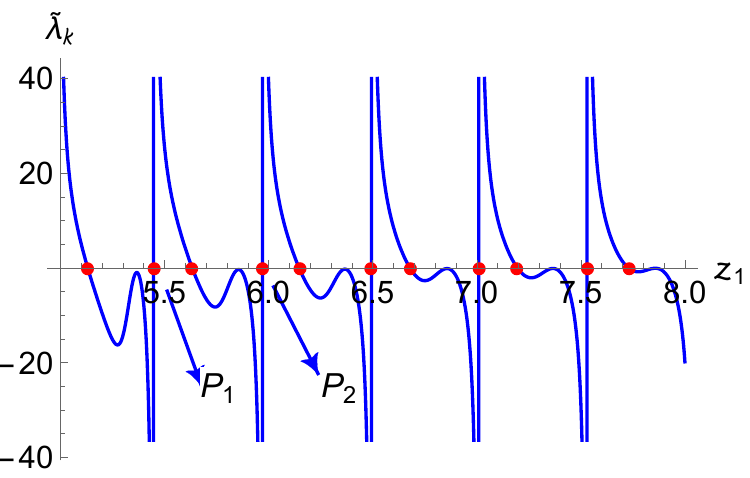}
\caption{\textit{ Behaviour of Modified Lyapunov exponent(LE) $\tilde \lambda_k$ for $k=m^{(1)}_{\rm eff}$}}
\label{LE}
\end{center}
\end{figure}

For a quantum chaotic system the OTOC measured in terms of the function $\mathcal{C}(t)$ is expected to behave as $\mathcal{C}(t)\sim e^{2\lambda t}$ with $\lambda$ being the $quantum$ $Lyapunov$ $exponent$. Therefore, the slope of $(1/2)$ln(OTOC) in time will give us the information of Lyapunov exponent $\lambda$. In our context we have been using dimensionless time-variable $z_1={m^{(1)}_{\rm eff}t}/{2\pi}$, and the associated dimensionless Lyapunov exponent will be $\tilde \lambda\equiv {2\pi \lambda}/{m^{(1)}_{\rm eff}}$ which implies $\mathcal{C}(z_1)\sim e^{2 \tilde{  \lambda}z_1}$. In Fig.\ref{lnOTOCbetaneq1qua},  we plotted $(1/2)$ln(OTOC) vs $z_1$, and linear growth in time can be clearly observed for a finite period of time. Therefore, the initial chaotic nature is manifested through the linear logarithmic growth of OTOC of a quantum field mode which is under the periodic driving force. 
Our chronological discussions so far seem to suggest that for a quantum chaotic system squeezing of quantum state parameter $r_k$, logarithmic growth of occupation number density ln($n_k$) and logarithm of highest OTOC amplitude have some intriguing relation, which we will discuss later. Furthermore, as expected chaos remains present in the system until the resonance ends or saturation begins which essentially hints probably the obvious fact that chaos and instability are intimately tied with each other \cite{Joras:2001yh}. Now our task is to determine the LE corresponding to different long wavelength modes. As one observes from Fig.\ref{lnOTOCbetaneq1qua},  growth of $\frac{1}{2}$ ln(OTOC) can be fitted by a straight line. We have the expression $\mathcal{C}(z_1)=A e^{2\tilde\lambda z_1}$ such that 
\begin{equation}\label{LEeq}
    \text{ln}~\mathcal{C}(z_1)=\text{ln}A+2\tilde\lambda z_1 \Rightarrow \frac{1}{2} \frac{d}{d z_1} \text{ln}\mathcal{C}(z_1)= \tilde \lambda ,
\end{equation}
where $A$ is any proportionality constant. For a particular $k$ mode the $\tilde\lambda_k=\frac{2\pi \lambda_k}{m^{(1)}_{eff}}$ in above expression (\ref{LEeq}) has many different zeros (\textcolor{red}{red} colored points) as in FIG. \ref{LE}. For a particular $k$ mode growth of OTOC depends upon a particular value of the Lyapunov exponent. However, Fig.\ref{LE} shows a time-varying nature LE, and we need to determine an effective value of this LE. To do so we prescribe the following procedure: We numerically compute the oscillation average of the function between two alternative zeros as marked by point $\text{P}_1$ at $z_1=5.5$ and $\text{P}_2$ at $z_1=6$ in Fig.\ref{LE}. Repeating this process from the initial time when the growth of a mode starts to the time when it saturates, we define the averaged effective LE $(\tilde{\lambda}^{eff}_k)$ as,
\[ \tilde{\lambda}_k^{eff} = \frac{\sum_{\rm oscillation}\langle \frac{1}{2} \frac{d}{d z_1} \text{ln} \mathcal{C}(z_1)\rangle|_{\rm oscillations}}{\text{Number of oscillation}} \] 
Using this procedure, we compute effective LE for various $k$ modes fixing the coupling strength $g$ and also for different coupling parameters $g$ with fixed $k$ mode (see Table \ref{tabLEneq1qua}). Finally in Fig.\ref{LEneq123qua} we have given pictorial representation of \textit{Lyapunov spectra}, that is $\tilde{\lambda}^{\rm eff}_k$ vs $\tilde{k}\equiv\big(k/m^{(1)}_{\rm eff}\big)$ and $\tilde{\lambda}^{\rm eff}_k$ vs $g$. From the figure, it is evident that the variation of LE with ${k}/{m^{(1)}_{\rm eff}}$ is not monotonic but rather chaotic. In spite of having this chaotic behavior, effectively with increasing $\tilde{k}$ resonance effect decreases as expected. On the other hand in coupling space, there seems to exist an effective critical coupling around which the system becomes maximally chaotic.
\begin{table}[ht]
\centering
\caption{\textit{Variation of effective LE$(\tilde{\lambda}_k^{\text{eff}})$ with $\tilde{k}$ and coupling strength \enquote{$g$} for $n=1 (g^2\phi^2\chi^2)$}}
\renewcommand{\arraystretch}{1.8}
\begin{tabular}{|c|c|c|c|c|c|c|c|c|c|c|c|}
\hline
\multirow{2}{*}{$g=5\times 10^{-4}$}  & $\tilde{k}$ & 1 &1.5  & 2 & 2.5& 3&3.5& 4 & 4.5 & 5& 6 \\ \cline{2-12} 
 & $\tilde{\lambda}_k^{\text{eff}}$ &0.919  &0.994  & 0.895 &0.586 & 1.037& 1.043 & 0.845 & 0.713 & 0.592 & 0.497    \\  
\hline
\multirow{2}{*}{$g=3\times 10^{-4}$}  & $\tilde{k}$ & 1 & 1.5 & 2 & 2.5 & 3 & 3.5 & 4 & 4.5 & 5 & 6 \\ \cline{2-12} 
&  $\tilde{\lambda}_k^{\text{eff}}$ & 0.475 & 0.712 & 0.704 & 0.930 & 0.575 & 0.690 & 0.601 & 0.411 & 0.406 & 0.201  \\ 
 \hline
 \multirow{2}{*}{$\tilde{k}=1$}  & $g\times 10^{-4}$ & 1 & 2 & 3 & 4 & 5 & 6 & 7 & 8 & 9 & 10 \\ \cline{2-12} 
 
 & $\tilde{\lambda}_k^{\text{eff}}$ &0.457  &0.468  & 0.475 &0.894 & 0.919& 1.190 & 1.186 & 1.214 & 1.153 & 1.133  \\  
\hline
 \multirow{2}{*}{$\tilde{k}=3$}  & $g\times 10^{-4}$ & 1 & 2 & 3 & 4 & 5 & 6 & 7 & 8 & 9 & 10 \\ \cline{2-12} 
 
 & $\tilde{\lambda}_k^{\text{eff}}$ &0.389  &0.454  & 0.575 &0.827 & 1.037& 0.887 & 0.916 & 1.055 & 1.086 & 1.090  \\  
\hline
\end{tabular}
\label{tabLEneq1qua}
\end{table}

\subsection{For $n=2$ :}
\begin{figure}[t]
\begin{center}
\includegraphics[scale=0.55]{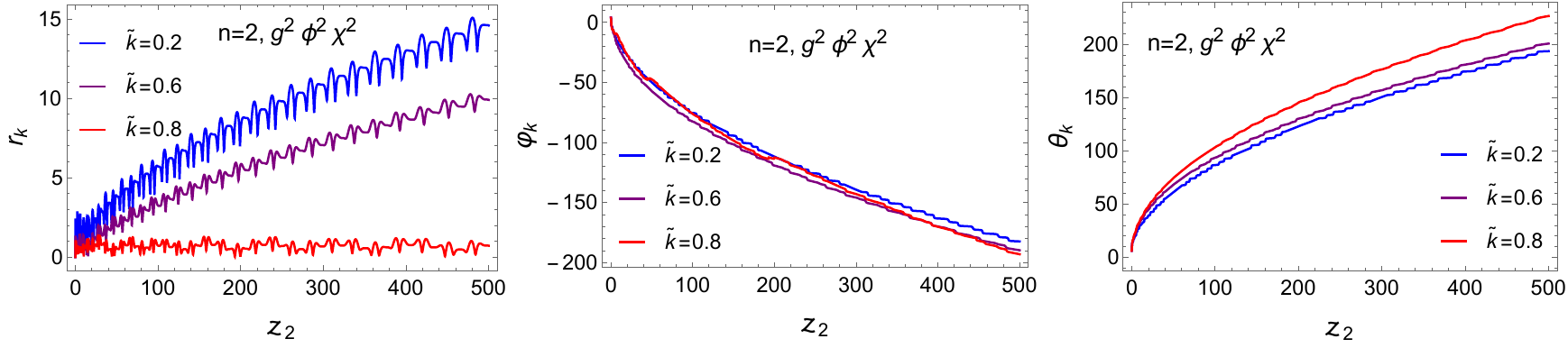}
\caption{\textit{Figure represents the time evolution of squeezing parameter$(r_k)$, squeezing angle$(\varphi_k)$ and rotation angle$(\theta_k)$ for three efficient $k$ modes, and the coupling strength is chosen to be $g=4\times 10^{-5}$. Here Squeezing angle$(\varphi_k)$ and rotation angle $(\theta_k)$ are given in unit of radian.}}
\label{sqzparaneq2qua}
\end{center}
\end{figure}

We have explicitly shown all the necessary steps of our study in the previous case for $n=1$ model. We present the numerical solution of $r_k, \varphi_k, \theta_k$ using the equations (\ref{3.29}) in the Fig.\ref{sqzparaneq2qua}.  
Unlike the model $n=1$, here squeezing parameter shows continuous growth over time without any late time saturation. As the particle number density is tied with the behavior of squeezing parameter $r_k$, the very nature of $r_k$ will have an impact on $n_k$ and this essentially causes an uninterrupted growth of $n_k$ over time.

This feature is special for $n=2$, when the inflaton field behaves like a relativistic fluid $w_\phi=1/3$ rendering vanishing Ricci scalar. Due to this simple fact, background inflaton field 
can be made to satisfy $\Psi^{''}+\bar{\lambda}\Psi^3=0$ \cite{Greene:1997fu} in conformal coordinate, which is independent of expansion. The solution of this equation is an oscillatory elliptic cosine function with constant amplitude. 
Such a feature is responsible for the uninterrupted growth of particle number over time for $n=2$ model as opposed to other $n$ values. In reality, such an uninterrupted growth process can be terminated by taking into account the backreaction of the produced particle into the background (for detail see \cite{Greene:1997fu,Maity:2018qhi}. In our present study, we are not incorporating those additional effects. For this case, therefore, in a similar manner we evaluate the Eq.(\ref{4.12}) and substituting the obtained results in (\ref{symplectic}) we present the behavior of OTOC amplitude in Fig.\ref{lnOTOCneq23qua}.
We further computed $\tilde{\lambda}_k^{\rm eff}$ with $\tilde{k}$ and coupling strength $g$. Taking various $k$ modes and coupling strength $g$ given in the Table \ref{tabLEneq2qua}, in Fig.\ref{LEneq123qua}, we show the variation of approximate effective LE with dimensionless momentum mode $\tilde{k}\equiv\big(k/m^{(2)}_{\rm eff}\big)$ and coupling strength $g$. Effective LE takes a prominent peak at some value of $\tilde{k}$ for a given coupling strength, then it gradually decreases with the increase of $\tilde{k}$. This sharp peak indicates the presence of a particular mode, for which the system is maximally chaotic. For $n=1$ model, this peak is less sharp compared to $n=2$. Variation of effective LE with $g$ follows the more or less same behavior as it is with $\tilde{k}$. For a given $\tilde{k}$, system peaks at a particular coupling, then it falls with the increase of $g$ with some fluctuations.     
\begin{table}[ht]
\centering
\caption{\textit{Variation of effective LE$(\tilde{\lambda}_k^{\text{eff}})$ with $\tilde{k}$ and coupling strength \enquote{$g$} for $n=2 (g^2\phi^2\chi^2)$}}
\renewcommand{\arraystretch}{1.8}
\begin{tabular}{|c|c|c|c|c|c|c|c|c|c|c|c|}
\hline
\multirow{2}{*}{$g=5.5\times10^{-5}$}  & $\tilde{k}$ & 0.1 &0.11  & 0.15 & 0.2& 0.25&0.3& 0.35 & 0.4 & 0.45& 0.5 \\ \cline{2-12} 
 & $\tilde{\lambda}_k^{\text{eff}}$ &0.421  &0.490  & 0.432 &0.420 & 0.394& 0.369 & 0.319 & 0.258 & 0.195 & 0.152    \\  
\hline
\multirow{2}{*}{$g=3.9\times 10^{-5}$}  & $\tilde{k}$ & 0.1 & 0.15 & 0.2 & 0.25 & 0.3 & 0.35 & 0.4 & 0.5 & 0.55 & 0.6 \\ \cline{2-12} 
&  $\tilde{\lambda}_k^{\text{eff}}$ & 0.319 & 0.381 & 0.534 & 0.600 & 0.626 & 0.493 & 0.549 & 0.609 & 0.932 & 0.258  \\ 
 \hline
 \multirow{2}{*}{$\tilde{k}=0.1$}  & $g\times 10^{-5}$ & 1.6 & 1.9 & 2.4 & 3.2 & 3.9 & 4.7 & 5.5 & 6.3 & 7.1& 7.9 \\ \cline{2-12} 
 
 & $\tilde{\lambda}_k^{\text{eff}}$ &0.358  &1.228  & 1.007 &0.507 & 0.319& 0.324 & 0.421 & 0.355 & 0.258 & 0.374  \\  
\hline
 \multirow{2}{*}{$\tilde{k}=0.3$}  & $g\times 10^{-5}$ & 1.6 & 1.9 & 2.4 & 3.2 & 3.9 & 4.7 & 5.5 & 6.3 & 7.1& 7.9 \\ \cline{2-12} 
 
 & $\tilde{\lambda}_k^{\text{eff}}$ &0.841  &0.974  & 0.295 &0.416 & 0.626& 0.316 & 0.369 & 0.349 & 0.290 & 0.286  \\  
\hline
\end{tabular}
\label{tabLEneq2qua}
\end{table}

\subsection{For $n=3$ :}
\begin{figure}[t]
\begin{center}
\includegraphics[scale=0.55]{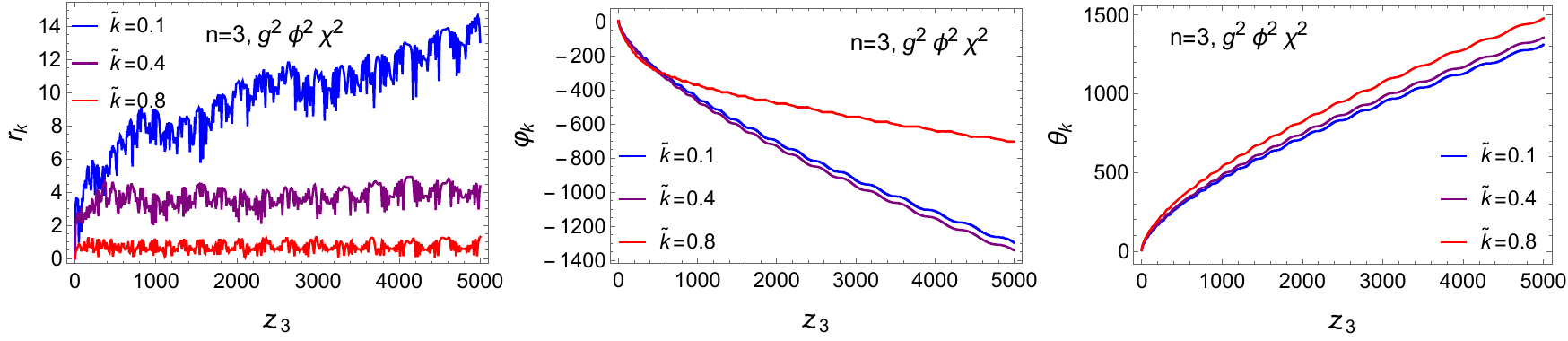}
\caption{\textit{Figure represents the time evolution of squeezing parameter$(r_k)$, squeezing angle$(\varphi_k)$ and rotation angle$(\theta_k)$ for three $k$ modes corresponding to significant production, and the coupling strength is chosen to be $g=4\times 10^{-5}$. Here Squeezing angle($\varphi_k$) and rotation angle ($\theta_k$) are given in units of radian.}}
\label{sqzparaneq3qua}
\end{center}
\end{figure}

\begin{figure}[t]
\begin{center}
\includegraphics[scale=0.50]{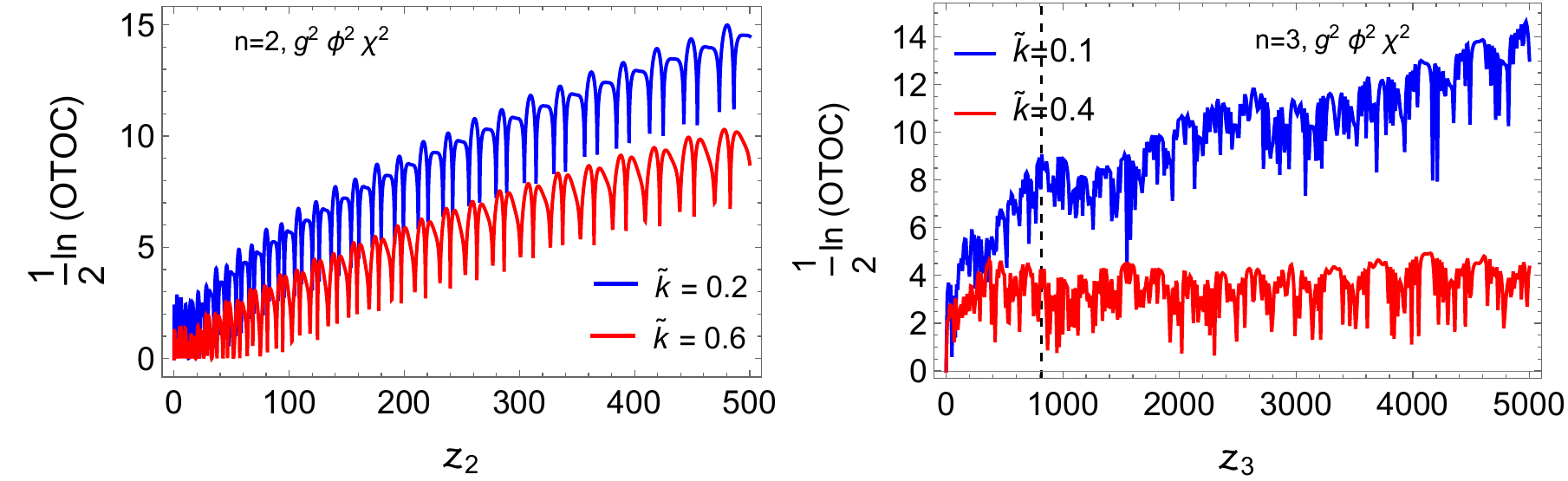} 
\caption{\textit{Figure represents the time variation of logarithm of OTOC amplitude for $n=2,3$ corresponding to two different $k$ modes with coupling strength $g=4\times10^{-5}$ for both cases. }}
\label{lnOTOCneq23qua}
\end{center}
\end{figure}

This particular value of $n$ induces a stiff fluid of inflaton with an equation of state $w_{\phi}=0.5$. In this background, the fluctuations evolve in a somewhat different manner compared to the models $n=1,2$. The behavior of squeezing parameters is depicted in Fig.\ref{sqzparaneq3qua}.
Here we observe that the growth time scale of squeezing parameter $r_k$ is very large and occurs over a very large number of background oscillations as opposed to the previous cases. In Fig.\ref{phineq123}, it is clearly seen that with the increase of background EoS, the decay rate of inflaton amplitude or in other words, the decay rate of the strength of the driving source gradually falls causing a slow energy transfer from background to the produced fluctuation. As long as the decay of inflaton amplitude takes place only because of background expansion, this slow energy transfer will essentially cause $r_k$ amplitude as well as OTOC amplitude(see Fig.\ref{lnOTOCneq23qua}) to grow very slowly over a long time.

\begin{figure}[t]
 \begin{center}
\includegraphics[scale=0.3]{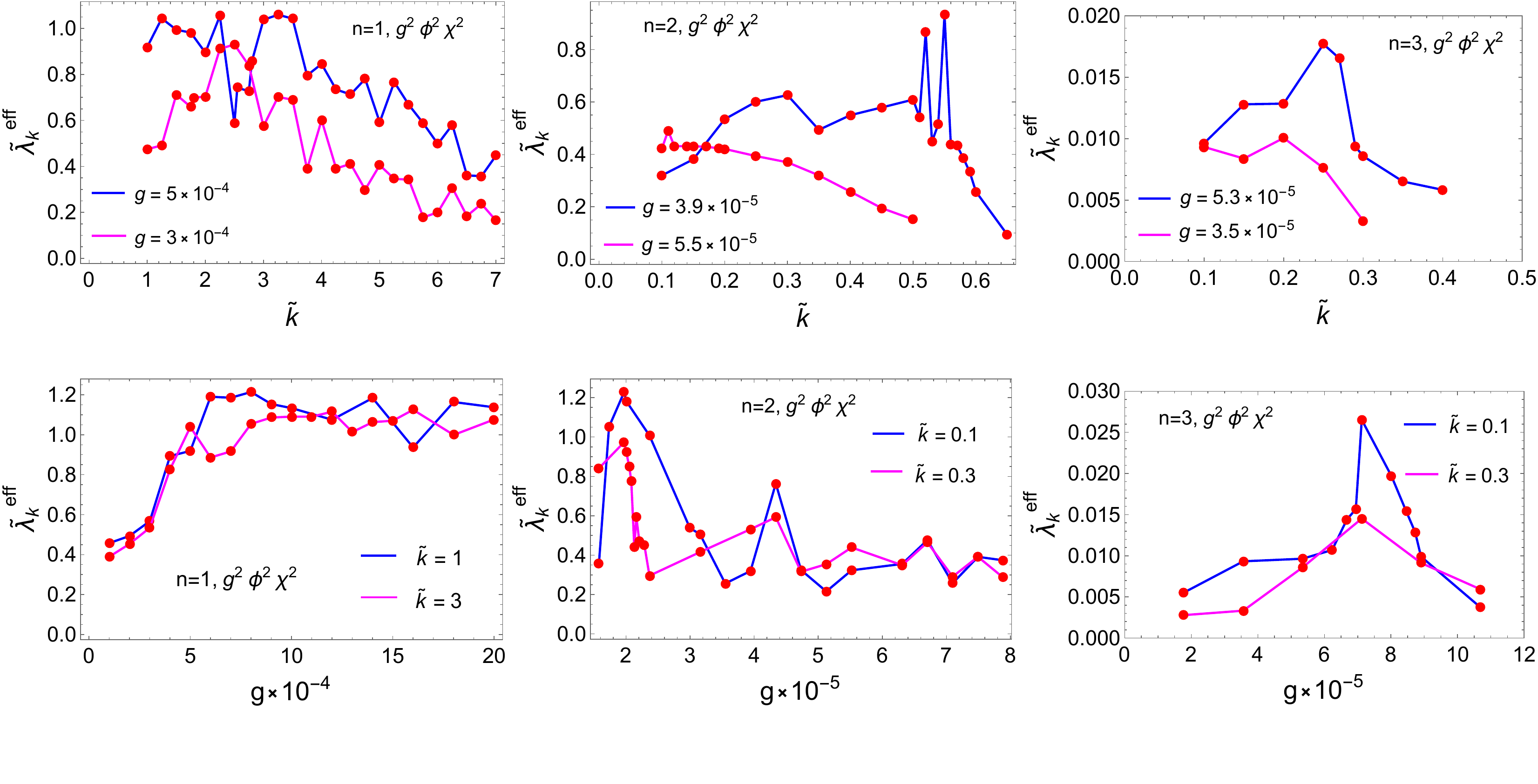}
\caption{\textit{Variation of $\tilde{\lambda}_k^{\rm eff}$ with $\tilde{k}\equiv (k/m^{(n)}_{\rm eff})$ and coupling strength }$g$ for three models.}
\label{LEneq123qua}
\end{center}
\end{figure}

Likewise, $n=2$ case, the absence of any saturation of OTOC amplitude within a finite time makes the determination of exact $\tilde{\lambda}_k^{\rm eff}$ for various $\tilde{k}$ and coupling $g$ a bit difficult. In order to have a rough estimate of the growth index of chaos, $\tilde{\lambda}_k^{\rm eff}$, for different $\tilde{k}$ and $g$, we take the first prominent peak just before a plateau type region. For example, in Fig.\ref{lnOTOCneq23qua}, we take first peak around $z_3\sim 820$ for $\tilde{k}=0.1$( See vertical dashed line in the right figure) and around $z_3\sim 380$ for $\tilde{k}=0.4$. Sticking to this idea of local saturation, we have given approximate numerical values of $\tilde{\lambda}_k^{\rm eff}$ corresponding to a few resonant modes for two different couplings in Table \ref{tabLEneq3qua}. We expect that the nature of the variation of $\tilde{\lambda}_k^{\rm eff}$ with two parameters($\tilde{k},g$) will remain unaffected by the choice of the aforesaid time scale. Stochastic nature, being an inherent property of the system results in a non-monotonic variation of $\tilde{\lambda}_k^{\rm eff}$ with $\tilde{k}$ and $g$ as seen in Fig.\ref{LEneq123qua}.

\begin{table}[ht]
\centering
\caption{\textit{Variation of effective LE$(\tilde{\lambda}_k^{\text{eff}})$ with $\tilde{k}$ and coupling strength \enquote{$g$} for $n=3 (g^2\phi^2\chi^2)$}}
\renewcommand{\arraystretch}{1.7}
\begin{tabular}{|c|c|c|c|c|c|c|}
\hline
\multirow{2}{*}{$g=5.3\times 10^{-5}$} & $\tilde{k}$ & 0.1 &0.15  & 0.2 & 0.25& 0.3  \\ \cline{2-7} 
 & $\tilde{\lambda}_k^{\text{eff}}$ &$9.639\times 10^{-3} $ & $12.798\times 10^{-3} $  & $ 12.854\times10^{-3} $ & $17.701\times 10^{-3}$ & $ 8.574\times 10^{-3} $  \\  
\hline
\multirow{2}{*}{$g=3.5\times 10^{-5}$}  & $\tilde{k}$ & 0.1 & 0.15 & 0.2 & 0.25 & 0.3   \\ \cline{2-7} 
&  $\tilde{\lambda}_k^{\text{eff}}$ & $9.324\times 10^{-3} $ & $8.380\times 10^{-3}$ & $10.059\times 10^{-3} $ & $7.609\times 10^{-3} $&  $3.324\times 10^{-3}$   \\ 
 \hline
 \multirow{2}{*}{$\tilde{k}=0.1$}  & $g\times 10^{-5}$ & 1.8  & 3.5  & 5.3 & 7.1 & 8.9 \\ \cline{2-7} 
 
 & $\tilde{\lambda}_k^{\text{eff}}$   & $5.530\times 10^{-3} $  & $9.324\times 10^{-3} $ & $9.639\times 10^{-3}$ & $26.524\times 10^{-3}$ & $9.899\times 10^{-3} $   \\  
\hline
 \multirow{2}{*}{$\tilde{k}=0.3$}  & $g\times 10^{-5}$ & 1.8  & 3.5  & 5.3 & 7.1 & 8.9   \\ \cline{2-7} 
 
 & $\tilde{\lambda}_k^{\text{eff}}$ & $2.786\times 10^{-3} $ & $3.324\times 10^{-3} $ & $8.574\times 10^{-3} $ & $14.517\times 10^{-3}$ & $9.127\times 10^{-2} $ \\  
\hline
\end{tabular}
\label{tabLEneq3qua}
\end{table}


\section{Squeezing and OTOC: for three-leg interaction ($\sigma\phi\chi^2)$ :}\label{sec9}

In this section, we shall repeat the same discussion as given in Section \ref{sec8} for three different $n=1,2,3$ taking three-leg interaction $\sigma\phi\chi^2$. Before presenting our numerical results for different models subject to this interaction, we need to give a short discussion on the parametric resonance process for this particular interaction. The structure of the resonance greatly relies on the nature of the interaction. To study the effect of tri-linear interaction on resonance process, the relevant part in the time-dependent frequency of massless case $\Omega_k^2(t)=\big(\frac{k^2}{a^2}+\sigma\phi-\frac{9}{4}H^2-\frac{3}{2}\dot H\big)$ (see Eq.\ref{3.5}) would be the first two terms. From the evolution of background inflaton in Fig.\ref{phineq123}, it is evident that when $\phi<0$ during one half of each oscillation, modes satisfying $k^2<\sigma|\phi(t)|a^2$ leads to negative $\Omega_k^2(t)$. In that case, the well-known $Tachyonic$ $instability$ sets in, and the modes experience exponential amplification. Therefore, for linear periodic driving force, the system encounters parametric as well as tachyonic instability which is a distinctive feature of tri-linear interaction. This process is known as $Tachyonic$ $resonance$\cite{Dufaux:2006ee,Felder:2001kt,Abolhasani:2009nb}. Both the instability leads to a very efficient production of fluctuation which causes the preheating phase to end within a few background oscillation. In the later discussion of this section, we will come across this distinctive feature of this particular interaction very closely. 

\begin{figure}[t]
    \begin{center}
\includegraphics[scale=0.50]{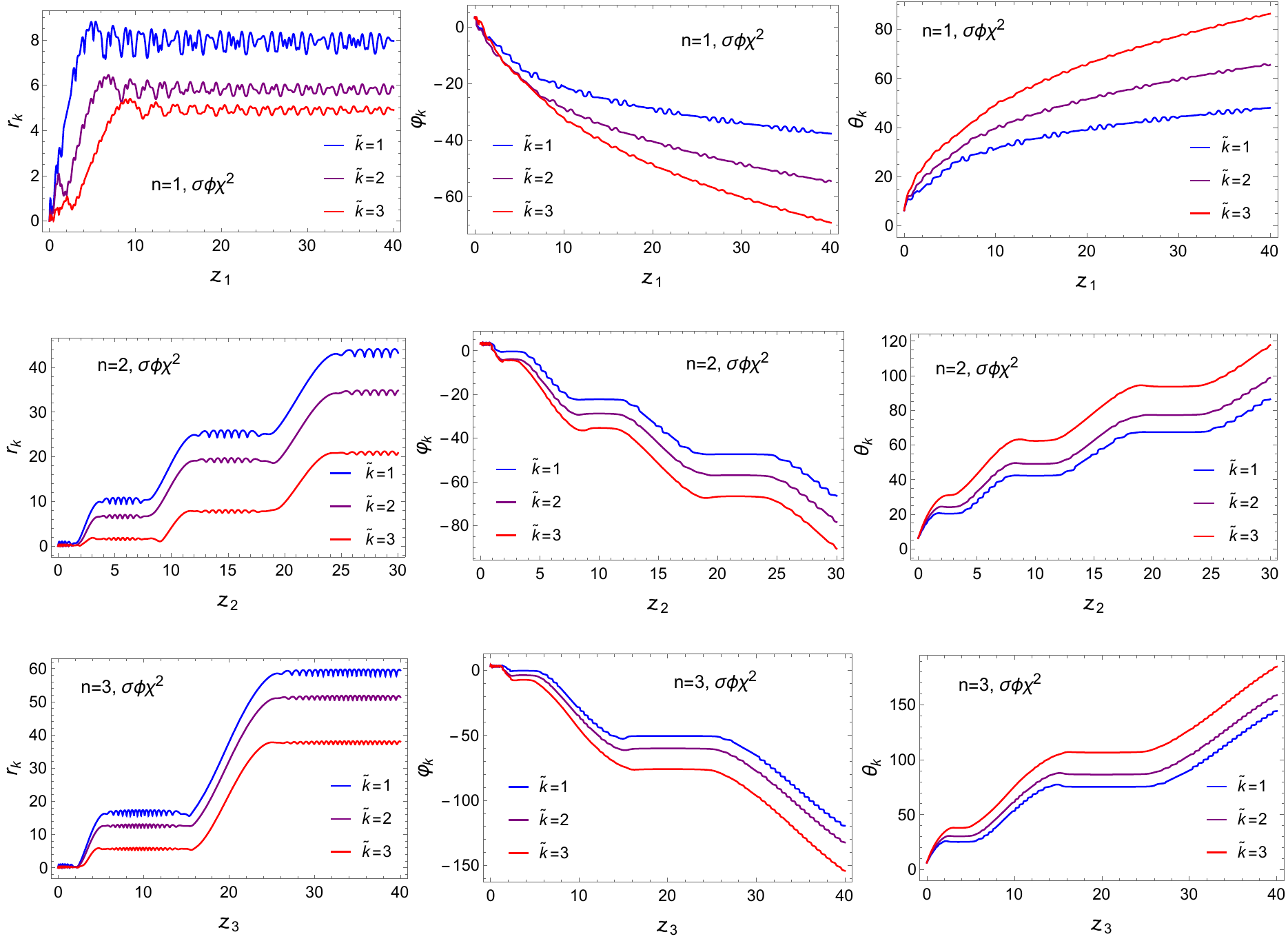}
\caption{\textit{Figure represents the time evolution of squeezing parameter$(r_k)$, squeezing angle$(\varphi_k)$ and rotation angle$(\theta_k)$ for $\tilde{k}= 1,2,3$ modes with dimensionless coupling strength $\tilde{g}=8\times 10^{-5}$ (first row), $\tilde{g}=9.8\times 10^{-5}$ (second row), and 
$\tilde{g}=1.1\cross 10^{-4}$ (third row). This parameter choice leads to parametric resonance in the system.}}
\label{sqzparaneq123tri}
\end{center}
\end{figure}

\begin{figure}[!ht]
\begin{center}
\includegraphics[scale=0.43]{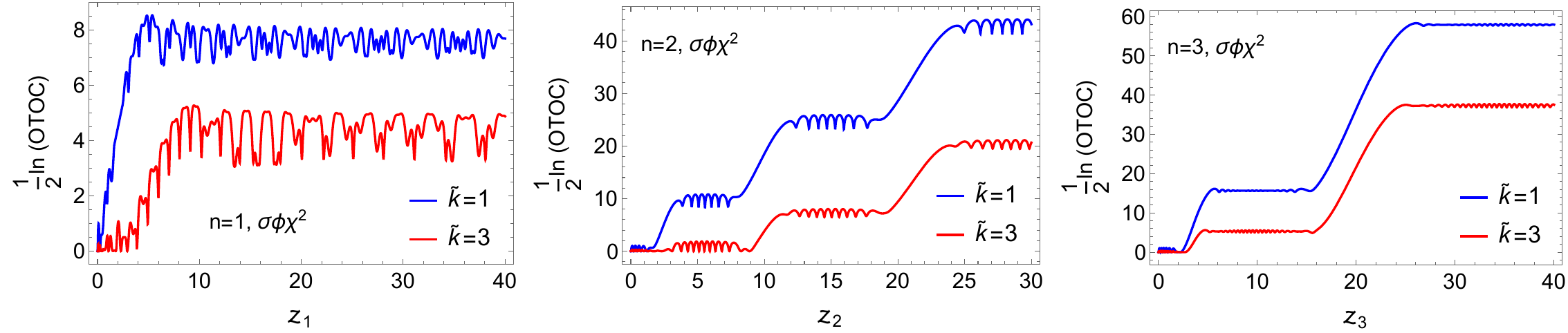} 
\caption{\textit{Figure represents the time variation of logarithm of OTOC amplitude for two efficient $\tilde{k} = 1,2$ modes with dimensionless coupling strength $\tilde{g}=8\times 10^{-5}$ (left panel), $\tilde{g}=9.8\times 10^{-5}$ (middle panel), and $\tilde{g}=1.1\times 10^{-4}$ (right panel). }}
\label{lnOTOCneq123tri}
\end{center}
\end{figure}

\begin{figure}[t]
    \begin{center}
\includegraphics[scale=0.29]{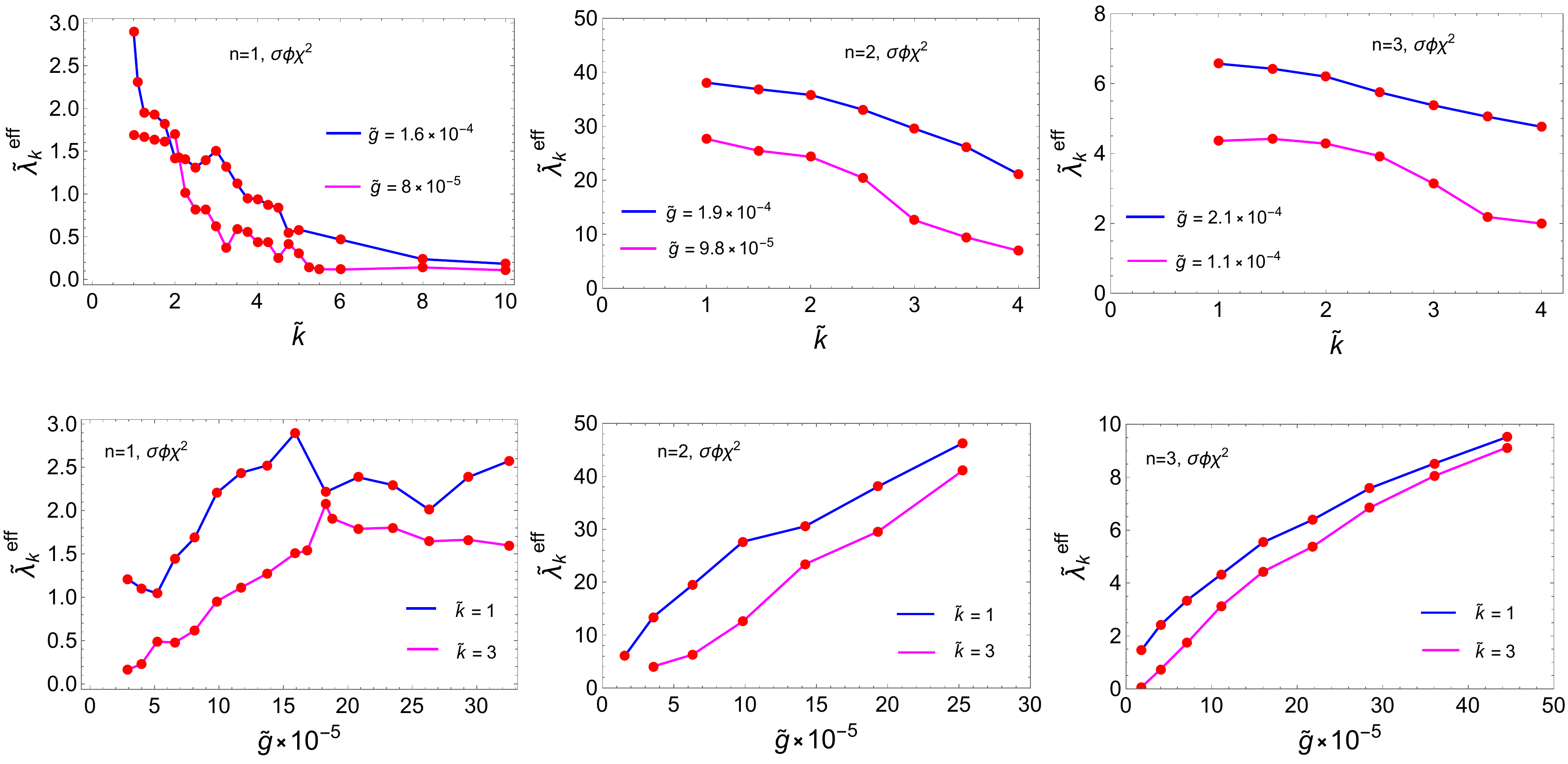}
\caption{\textit{ Figure represents the Variation of $\tilde{\lambda}_k^{\text{eff}}$ with dimensionless modes $\tilde{k}$ and dimensionless coupling $\tilde{g}$ for three models in three-leg type interaction. }}
\label{LEneq123tri}
\end{center}
\end{figure}

Nevertheless, we will follow the same analysis and the behavior of the three squeezing parameters is depicted in Fig.\ref{sqzparaneq123tri} for different values of $n$.
Both qualitative and quantitative differences can be observed in the smooth nature of the growth of squeezing and OTOC along with its amplitude.
For $n=1$, saturation time scale turns out to be for $\tilde{k}=1$,  $z_1\sim 5.2$, for $\tilde{k}=2$, $z_1\sim 6.9$ and for $\tilde{k}=3$, $z_1\sim 9.2$.
For $n=2$, due to effective flatness perceived by both background and fluctuations, every mode is expected to have eternal growth in $r_k$. 
For all the cases the growth seems to be less stochastic compared to the four-leg interaction. This property will make an impact on the chaotic nature of the system. Using Eq.(\ref{symplectic}), we also plotted the nature of OTOC to capture the chaotic dynamics of the system for three models(See Fig.\ref{lnOTOCneq123tri}).  
In Fig.\ref{lnOTOCneq123tri}, for $n=1$ model, we notice that for $\tilde{k}=1$, OTOC amplitude peaks around $z_1\sim 5.2$ and $z_1\sim 9.2$ for $\tilde{k}=3$ and then it fluctuates without further growth. Finally employing the same methodology that was described in the last Section \ref{sec8}, we estimate the effective LE $\tilde\lambda^{\rm eff}_k$ for different values of $k$ with fixed $g$, and for different $g$ with fixed $k$ (see Fig.\ref{LEneq123tri}). In the Table \ref{tabLEneq1tri}, Table \ref{tabLEneq2tri} and Table \ref{tabLEneq3tri}, we have given some numerical values of $\tilde{\lambda}^{\rm eff}_k$ for different momentum modes $\tilde{k}$ and coupling constant $\tilde{g}\equiv \big(\sigma/m^{(n)}_{\rm eff}\big)$ for three different models.

\begin{table}[ht]
\centering
\caption{\textit{Variation of effective LE$(\tilde{\lambda}_k^{\rm eff})$ with $\tilde{k}$ and dimensionless coupling strength \enquote{$\tilde{g}$} for $n=1 (\sigma\phi\chi^2)$}}
\renewcommand{\arraystretch}{1.8}
\begin{tabular}{|c|c|c|c|c|c|c|c|c|c|c|c|}
\hline
\multirow{2}{*}{$\tilde{g}=1.6\times10^{-4}$}  & $\tilde{k}$ & 1 &1.5  & 2 & 2.5& 3&3.5& 4 & 4.5 & 5& 6 \\ \cline{2-12} 
 & $\tilde{\lambda}_k^{\text{eff}}$ &2.896  &1.928  & 1.421 &1.312 & 1.507& 1.117 & 0.943 & 0.837 & 0.583 & 0.466    \\  
\hline
\multirow{2}{*}{$\tilde{g}=8\times 10^{-5}$}  & $\tilde{k}$ & 1 & 1.5 & 2 & 2.5 & 3 & 3.5 & 4 & 4.5 & 5 & 6 \\ \cline{2-12} 
&  $\tilde{\lambda}_k^{\text{eff}}$ & 1.685 & 1.635 & 1.697 & 0.813 & 0.617 & 0.594 & 0.435 & 0.251 & 0.302 & 0.115  \\ 
 \hline
 \multirow{2}{*}{$\tilde{k}=1$}  & $\tilde{g}\times 10^{-5}$ & 2.9 & 3.9 & 5.2 & 8 & 9.8 & 11.7 & 16 & 18.2 & 20.8 & 32.5 \\ \cline{2-12} 
 
 & $\tilde{\lambda}_k^{\text{eff}}$ &1.208  &1.104  & 1.044 &1.685 & 2.203& 2.435 & 2.896& 2.214 & 2.385 & 2.571  \\  
\hline
 \multirow{2}{*}{$\tilde{k}=3$}  & $\tilde{g}\times10^{-5}$ &  2.9 & 3.9 & 5.2 & 8 & 9.8 & 11.7 & 16 & 18.2 & 20.8 & 32.5  \\ \cline{2-12} 
 
 & $\tilde{\lambda}_k^{\text{eff}}$ &0.165  &0.227  & 0.488 &0.617 & 0.952& 1.109 & 1.507 & 2.072 & 1.789 & 1.598  \\  
\hline
\end{tabular}
\label{tabLEneq1tri}
\end{table}

\begin{table}[ht]
\centering
\caption{\textit{Variation of effective LE$(\tilde{\lambda}_k^{\rm eff})$ with $\tilde{k}$ and dimensionless coupling strength \enquote{$\tilde{g}$} for $n=2 (\sigma\phi\chi^2)$}}
\renewcommand{\arraystretch}{1.6}
\begin{tabular}{|c|c|c|c|c|c|c|c|}
\hline
\multirow{2}{*}{$\tilde{g}=1.9\times 10^{-4}$} & $\tilde{k}$ & 1 &1.5  & 2 & 2.5& 3& 4  \\ \cline{2-8} 
 & $\tilde{\lambda}_k^{\text{eff}}$ &38.066  &36.861  & 35.799 &33.094 & 29.545& 21.114    \\  
\hline
\multirow{2}{*}{$\tilde{g}=9.8\times 10^{-5}$}  & $\tilde{k}$ & 1 & 1.5 & 2 & 2.5 & 3  & 4  \\ \cline{2-8} 
&  $\tilde{\lambda}_k^{\text{eff}}$ & 27.649 & 25.468 & 24.339  & 20.486 & 12.602 & 6.946   \\ 
 \hline
 \multirow{2}{*}{$\tilde{k}=1$}  & $\tilde{g}\times 10^{-5}$  & 3.6 & 6.3 & 9.8 & 14.1 & 19.3 & 25.2  \\ \cline{2-8} 
 
 & $\tilde{\lambda}_k^{\text{eff}}$   &13.385  & 19.452 &27.649 & 30.519& 38.066 & 46.235   \\  
\hline
 \multirow{2}{*}{$\tilde{k}=3$} & $\tilde{g}\times 10^{-5}$  & 3.6 & 6.3 & 9.8 & 14.1 & 19.3 & 25.2  \\ \cline{2-8} 
 
 & $\tilde{\lambda}_k^{\text{eff}}$ &4.052  &6.270  & 12.602 &23.364 & 29.545& 41.041    \\  
\hline
\end{tabular}
\label{tabLEneq2tri}
\end{table}

\begin{table}[ht]
\centering
\caption{\textit{Variation of effective LE$(\tilde{\lambda}_k^{\rm eff})$ with $\tilde{k}$ and dimensionless coupling strength \enquote{$\tilde{g}$} for $n=3 (\sigma\phi\chi^2)$}}
\renewcommand{\arraystretch}{1.6}
\begin{tabular}{|c|c|c|c|c|c|c|c|c|}
\hline
\multirow{2}{*}{$\tilde{g}=2.1\times 10^{-4}$} & $\tilde{k}$ & 1 &1.5  & 2 & 2.5& 3&3.5& 4  \\ \cline{2-9} 
 & $\tilde{\lambda}_k^{\text{eff}}$ &6.567  &6.424  & 6.195 &5.746 & 5.374& 5.050 & 4.764    \\  
\hline
\multirow{2}{*}{$\tilde{g}=1.1\times 10^{-4}$}  & $\tilde{k}$ & 1 & 1.5 & 2 & 2.5 & 3 & 3.5 & 4  \\ \cline{2-9} 
&  $\tilde{\lambda}_k^{\text{eff}}$ & 4.360 & 4.419 & 4.280  & 3.926 & 3.134 & 2.184 & 1.996  \\ 
 \hline
 \multirow{2}{*}{$\tilde{k}=1$}  & $\tilde{g}\times 10^{-5}$ & 1.8 & 4 & 7.1 & 11.1 & 16 & 28.5 & 44.5  \\ \cline{2-9} 
 
 & $\tilde{\lambda}_k^{\text{eff}}$   &1.460  & 2.407 &3.347 & 4.360& 5.542 & 7.585 & 9.518   \\  
\hline
 \multirow{2}{*}{$\tilde{k}=3$}  & $\tilde{g}\times 10^{-5}$ & 1.8 & 4 & 7.1 & 11.1 & 16 & 28.5 & 44.5 \\ \cline{2-9} 
 
 & $\tilde{\lambda}_k^{\text{eff}}$ &0.057  &0.732  & 1.737 &3.134 & 4.433& 6.848 & 9.123   \\  
\hline
\end{tabular}
\label{tabLEneq3tri}
\end{table}

\section{Relation between Floquet exponent and Lyapunov exponent, and OTOC Spectrum}\label{sec10}
So far we mainly discussed the nature of OTOC and calculated the associated Lyapunov exponents. However,
it is understood that such growth is intimately connected with the 
resonant production of modes. We, therefore, expect to have a direct relation between the
Lyapunov and Floquet exponent which are the measure of chaos and resonance respectively under the influence of periodic driving force. To establish an approximate analytic relation between those two quantities, we shall resort to the method of successive scattering on parabolic potential discussed in detail\cite{Kofman:1994rk,Kofman:1997yn}. Here we quote the main results. If we investigate the resonance process more closely we notice that whenever the background inflaton crosses zero at the minimum, the particle number density shows a sharp growth. Otherwise, particle number density behaves as an adiabatic invariant quantity between two successive zero crossings. Such behavior allows one to obtain an approximate analytical expression of field mode in terms of Floquet exponent around each zero crossing time, say $j$-th crossing time is defined to be $t_j$. If one consider $\phi^2 \chi^2$ interaction, near $t_j$, massless field mode Eq.(\ref{3.5}) for $n=1$ assumes following form in the parabolic potential around $t=t_j$,  
\begin{equation}\label{flucteq1}
    \frac{d^2X_k}{dt^2}+\bigg(\frac{k^2}{a^2}+g^2\Phi_j^2 m^{(1)2}_{\rm eff}(t-t_j)^2\bigg)X_k=0
\end{equation}
where $(\Phi_j)$ is the time-dependent amplitude of the inflaton calculated at the instant of $j$-th crossing $t_j$. For $|t-t_j| >0$, the adiabaticity condition is satisfied. Using the Bogoliubov method  \cite{Kofman:1994rk,Kofman:1997yn},one can compute the rescaled Floquet exponent as
\begin{equation}\label{floqueteq}
    \tilde{\mu}^j_k =\ln \Big(1+2e^{-\pi\kappa_j^2}-2\sin \theta^{\text{tot}}_j e^{-\frac{\pi}{2}\kappa_j^2}\sqrt{1+e^{-\pi\kappa_j^2}}\Big) .
\end{equation}
Where, $\kappa_j^2={k^2}/({a(t_j)^2 gm^{(1)}_{\rm eff}\Phi_j})$, and $\theta^{\text{tot}}_j$ the total random phase accumulated by the field wave from the initial time to the instant $t_j$. In order to obtain the nature of OTOC spectra as well as the possible relation between Floquet and Lyapunov exponent, we need to evaluate first the commutation relation between field($X_k$) and conjugate momenta($\Pi_k$) as outlined in Section \ref{sec4}. 
We closely follow the reference \cite{Kofman:1997yn} in deriving the amplitude of the commutation relations using the asymptotic form of the field mode $X_k$ as in Eq.(\ref{wkb}). We write the expression of OTOC amplitude as follows:

\begin{align}\label{OTOCamp1}
   |X_k(t_j) \dot X_k^{\star}(t_0)-X_k^{\star}(t_j)\dot X_k(t_0)|^2 =\frac{e^{\tilde{\mu}^j_k.j}\Omega_k(t_0)(1+{\rm cos}\mathcal{V})^2}{2\Omega_k(t_j)}
\end{align}
Where $\mathcal{V}$ is a constant phase, $t_j=\frac{\pi j}{m_{\rm eff}^{(1)}}$ and $t_0$ is the initial time. Using this amplitude, the expression of OTOC spectra for $\phi^2\chi^2$ type interaction at any later time $t$ becomes
\begin{equation}\label{OTOCamp2}
    \text{ln(OTOC)}\equiv {\rm ln}\Big(1+2e^{-\pi\kappa^2}-2\text{sin}\theta^{\text{tot}} e^{-\frac{\pi}{2}\kappa^2}\sqrt{1+e^{-\pi\kappa^2}}\Big).\frac{m_{\rm eff}^{(1)}t}{\pi}+{\rm ln}\Big(\frac{\Omega_k(t_0)}{2\Omega_k(t)}\Big)+2  \ln (1+\cos \mathcal{V})
\end{equation}
We have used Eq.(\ref{floqueteq}) to reach the final form of the spectra in Eq.(\ref{OTOCamp2}).
Now associated with the Lyapunov exponent as outlined in section \ref{sec4} and using (\ref{wkb}) we write, Lyapunov exponent at the instant of $j$-th crossing as, 
\begin{align}\label{LEanalytic}
    & e^{2\lambda_kt_j }= 
     e^{\tilde{\mu}^j_k.j } \Big(\frac{\Omega_k(t_0)}{2\Omega_k(t_j)}\Big)(1+\text{cos}\mathcal{V})^2\nno\\
    &\Rightarrow \tilde{\lambda}^j_k=\tilde{\mu}^j_k+\frac{1}{j}{\rm ln}\Big(\frac{\Omega_k(t_0)}{2\Omega_k(t_j)}\Big)+\frac{2}{j}{\rm ln}(1+\text{cos}\mathcal{V})
\end{align}
Here the out-of-time order commutation of one combination field and conjugate momenta has been defined between $t_0$ and the $j$-th crossing instant $t_j$.\\ The above expression is true at the point of a particular zero crossing $t_j$. We, therefore, can have an effective relation 
\begin{equation}\label{LEfloquet}
     \tilde{\lambda}^{\rm eff}_k = \tilde{\mu}^{\rm eff}_k +  \overline{\frac{1}{j}\ln \Big({\frac{\Omega_k(t_0)}{2\Omega_k(t_j)}}\Big)} + 2~\overline{\frac{1}{j} \ln (1+\cos \mathcal{V})},
\end{equation}
where the average is taken over time from the initial to the saturation time scale. So practically we take average over the number of oscillations executed by the background within the time required for saturation of a particular mode. In the above expression, the phase $\mathcal{V}$ is random around $\pi/2$. So, after taking the average over the entire saturation time scale for a particular mode, the third term will have a vanishing contribution. For the second term, we have numerically found a smaller contribution compared to $\tilde{\mu}_k^{\rm eff}$. Finally, we find $\tilde{\lambda}_k^{\rm eff}$ and $\tilde{\mu}_k^{\rm eff}$ are more or less the same which is consistent with the numerical result. 

\begin{figure}[t]
    \begin{center}
\includegraphics[scale=0.28]{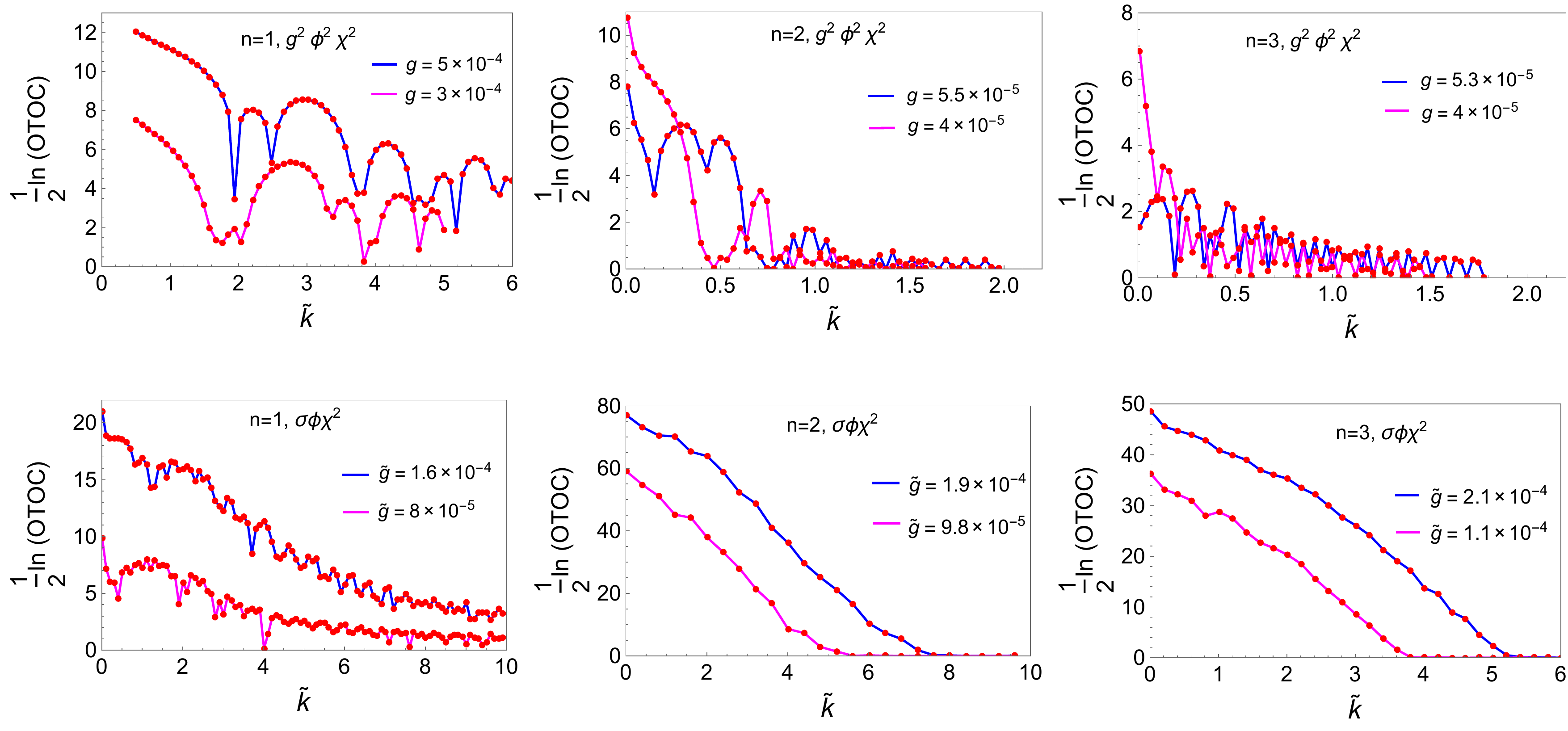}
\caption{\textit{Figure represents the spectral behaviour of} ln(OTOC) \textit{ with two different coupling strengths for two interactions. }}
\label{OTOCspectra}
\end{center}
\end{figure}

From Eq.(\ref{OTOCamp2}) it is clearly seen that the spectral nature of OTOC is governed by the nature of the Floquet exponent. The growth index of particle number in Eq.(\ref{floqueteq}) being a function of random phase in expanding background causes somewhat random growth of number density after every zero crossing or every half of a period. This non-monotonic behavior is also reflected in the spectral behavior of OTOC amplitude as shown in Fig.\ref{OTOCspectra}. The random oscillatory behavior of OTOC amplitude in momentum space is also caused by the randomness of the Floquet exponent in the context of stochastic resonance as discussed in detail in \cite{Kofman:1997yn}.  Out of three different background driving sources, only for $n=1$ the system reach a saturation state \footnote{Late time saturation of different modes can be assumed to be the reminiscent of thermalization of a typical thermodynamic system.} after going through chaotic dynamics within a finite time. While evaluating OTOC spectra numerically for $n=1$ model we have chosen the saturation time corresponding to the maximum mode that will be produced in a broad resonance regime. Unlike $n=1$ model, no such saturation is observed for $n=2$ model as shown in Fig.\ref{lnOTOCneq23qua} and Fig.\ref{lnOTOCneq123tri}(Reason is discussed in Section \ref{sec8}). For this case, we computed OTOC spectra by truncating the evolution at any arbitrary instant of time for each momentum mode. With the increase of potential exponent $n$, slow decay of the amplitude of inflaton causes slow growth of OTOC amplitude (See Fig.\ref{lnOTOCneq23qua} for $n=3$), and for this, we have considered a local saturation time corresponding to the maximum mode chosen in the spectra in Fig.\ref{OTOCspectra}, where the coupling parameters are chosen from the broad resonance regime (See the lower bound in Eq.(\ref{couplingbound})). Despite several fluctuations, one common feature observed in the spectra for all the models is that the higher the momentum, the lower the OTOC amplitude which actually mimics the nature of resonant particle production. The existence of different momentum bands for different values of resonance strength parameter ($q_g, q_{\sigma}$) can be observed to be nicely imprinted in the broad oscillatory feature in the OTOC spectrum, particularly for quadratic coupling. However, for tri-linear coupling due to tachyonic and broad resonance, such oscillatory nature smoothens out.    

\section{Squeezing, Chaos and thermalization}\label{sec12}

\subsection{Poincar$\acute{\text{e}}$ section: Semi-classical visualization}
\begin{figure}[t]
    \begin{center}
\includegraphics[scale=0.55]{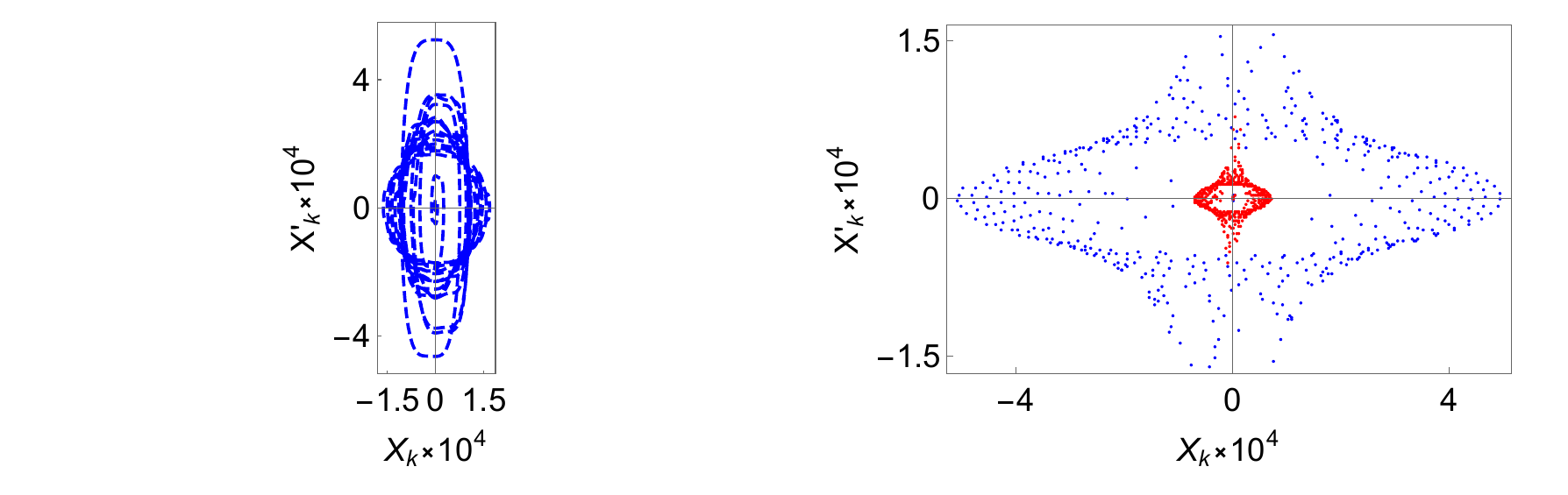}
\includegraphics[scale=0.605]{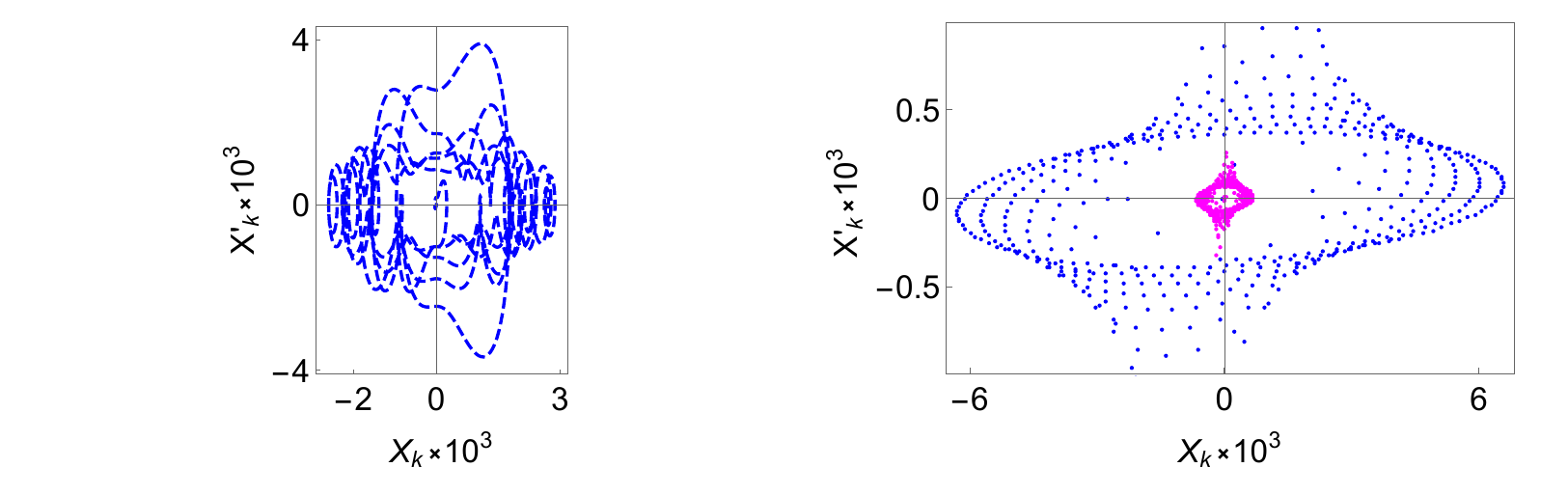} \caption{\textit{\textbf{Upper Panel:} Figure on the left represents the parametric plot of $X_k$ and $X_k^{'}$ where $\tilde{k}=1$ and figure on the right represents the Poincar$\acute{\text{e}}$ section for $n=1$ model for $\phi^2\chi^2$ interaction. Different colors correspond to different initial conditions. In the Poincar$\acute{\text{e}}$ section, \textcolor{blue}{Blue} colored points are phase space points for $\tilde{k}=1$ and \textcolor{red}{Red} colored points are phase space points for $\tilde{k}=3.$ Here dimensionless coupling strength $g=5\cross 10^{-4}$. \textbf{Lower Panel:} Figure on the left represents the parametric plot of $X_k$ and $X_k^{'}$ where $\tilde{k}=1$ and figure on the right represents the Poincar$\acute{\text{e}}$ section for $n=1$ model for $\sigma\phi\chi^2$ interaction. In the Poincar$\acute{\text{e}}$ section, \textcolor{blue}{Blue} colored points are phase space points for $\tilde{k}=1$ and \textcolor{magenta}{Magenta} colored points are phase space points for $\tilde{k}=2.$ Here dimensionless coupling strength $\tilde{g}=8\cross 10^{-5}$ . }}
\label{phasespace}
\end{center}
\end{figure}

At the outset of this section, we gain some insight into how the chaotic nature of the system is connected to the squeezing of states. It is known that positive LE signifies the chaotic nature of a system. It actually measures the sensitivity of the system to its initial condition. The well-known method to explore chaotic behavior is to generate and analyze the Poincar$\acute{\text{e}}$ sections, particularly for non-integrable systems. Each field mode of our present system is periodically driven one dimensional parametric oscillator which does not have any conserved quantity. For this system, we identified the semi-classical phase space variables by calculating their expectation values in the squeezed quantum state.   
We obtain the Poincar$\acute{\text{e}}$ section or a surface of section, by mapping those semi-classical phase space variables projecting onto two dimensions. We generate this Poincar$\acute{\text{e}}$ section by sampling the data points in a 2D plane at a regular interval set by the periodicity of the driving force namely the oscillatory inflaton field. 
We present Poincar$\acute{\text{e}}$ section for $n=1$ model for both types of interactions(three-leg and four-leg), besides this, we also give the parametric plot of $X_k$ and $X_k^{'}$ which carries the information of squeezing of the system for both models(See Fig.\ref{phasespace}).
So for a non-chaotic system, we expect that all the data points will lie in an orbit and will form closed orbits, but here we see that the phase space points are scattered in the plane and are forming a cloud of points, which signifies the presence of chaos in the system. 
This typical behavior also tells us about the squeezing of the system. Looking at Fig.\ref{sqzparaneq1qua} we readily understand that the squeezing parameter of this system corresponding to particular parameter values ($k,g$) grows over a certain duration and then gets saturated, that means after the point of saturation there is no further squeezing in the system. Furthermore, the growth of OTOC in Fig.\ref{lnOTOCbetaneq1qua} also exhibits the same behavior. Connecting these two outcomes together, we can state that as long as squeezing is there, the system will remain chaotic which vindicates the statement that \textit{highly squeezed states are prime for quantum chaos}.\\ 
In the following section, we would like to shed light on the possible thermalization temperature of such a chaotic system. 
\subsection{Defining thermalization temperature and Maldacena-Shenker-Stanford (MSS) bound}
\begin{figure}[t]
    \begin{center}
\includegraphics[scale=0.35]{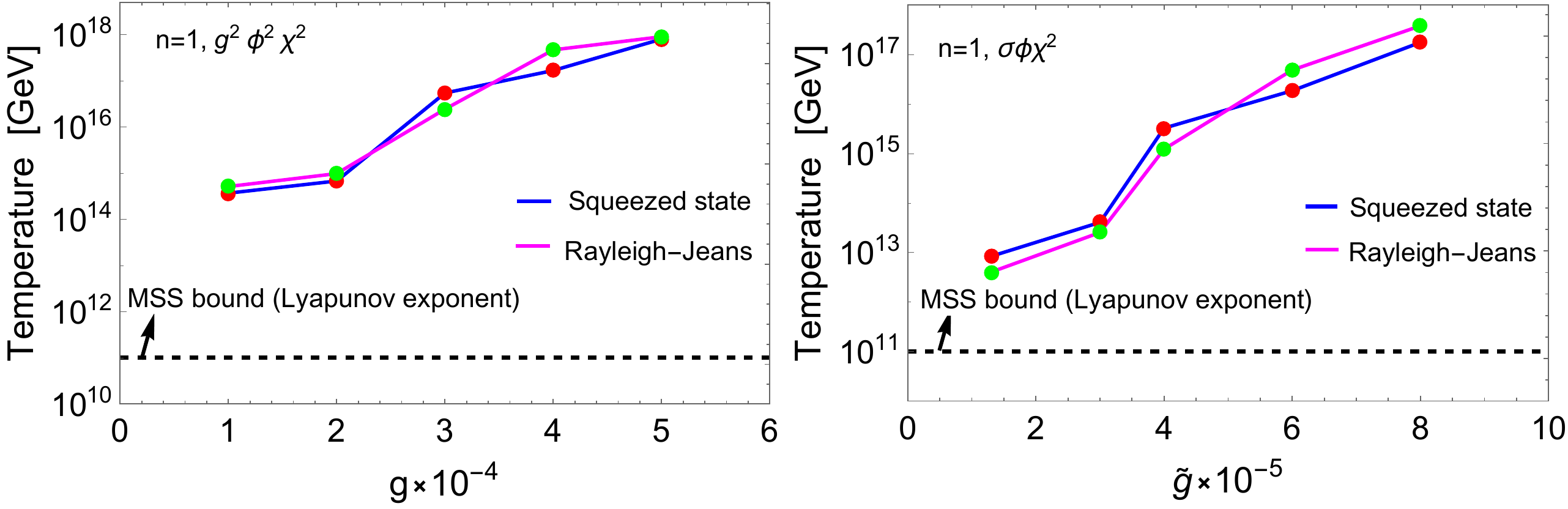}
\caption{\textbf{Left Panel:}\textit{ Figure represents the variation of effective temperature with coupling strength for four-leg type interaction.} \textbf{Right Panel:}\textit{Figure represents the variation of effective temperature with coupling strength for three-leg type interaction. It is seen that effective temperatures for different coupling are consistently following the lower bound of temperature for both interactions. Here \textcolor{red}{red} dots indicate the temperature obtained through the relation (\ref{inversetemp2}) and \textcolor{green} {green} dots indicate the temperature obtained through Rayleigh-Jeans formula given in (\ref{Rayleightemp}) for both the interactions.}}
\label{MSSbound}
\end{center}
\end{figure}

  {\bf MSS bound:} Chaos and thermalization have been the subject of investigation over a long period of time in non-equilibrium field theory. In this regard an intriguing bound on chaos has been proposed in \cite{Maldacena:2015waa} which we call MSS bound in brief, and the Lyapunov exponent is universally conjectured to be bounded as $\lambda \leq 2\pi (k_B/\hbar) T$, where $T$ is assumed to be the temperature of the chaotic system when it becomes thermalized after the relaxation from chaotic instability. What it suggests the growth of chaos of a thermal quantum system is bounded by the system temperature $T$. If we reverse this argument, one may arrive at the following universal lower bound on the temperature of a quantum chaotic system  $T \geq  (\hbar/(2 \pi k_B))\lambda$. In our present system, we apply the proposed bound and estimate the minimum possible temperature that the system can achieve. So far we have discussed the chaotic nature and calculated the value of the associated Lyapunov exponents of a quantum scalar field when coupled with the oscillating inflaton. Further, we have performed our analysis for individual momentum modes and computed the Lyapunov exponent for each mode. Therefore, instead of getting the information about the temperature of the entire system, we will have the information of each mode. Using the bound we may write, 
\begin{equation}\label{MSSbound1}
T_k \geq  \frac {\hbar}{2 \pi k_B}\lambda_k.
\end{equation}  
Where $\lambda_k$ and $T_k$ are LE and the corresponding minimum possible temperature perceived by the mode. A more rigorous estimation would be to do the analysis in real space by using lattice simulation technique and identifying MSS bound. We will defer the lattice studies for our future work. Using the approximate values of effective LE given in Table \ref{tabLEneq1qua} and Table \ref{tabLEneq1tri} for $\phi^2\chi^2$ and $\sigma\phi\chi^2$ interaction respectively, we can identify the lower bound of the temperature of the system by averaging over the resonant momentum band as follows,
\bea\label{MSSbound2}
{\bar T}_{\rm MSS} \geq \frac {\hbar}{2 \pi k_B} \frac {\int_{\rm band} \lambda_{k} k^2 dk}{\int_{\rm band}  k^2 dk}
\eea 
When the chaotic system reaches thermal equilibrium, the minimum temperature achieved by the system will be given by the following bound in Eq.(\ref{MSSbound2}). By exploiting this we obtain the lower bound of temperature for different coupling strengths for two different interactions. 
Using Eq.(\ref{MSSbound2}), for example, we found $\bar{T}_{\rm MSS}=2.62\times 10^{11}(4.76\times 10^{11})$ GeV for $g=3\times 10^{-4}(5\times10^{-4})$ in case of $g^2\phi^2\chi^2$ interaction  and $\bar{T}_{\rm MSS}=2.33\times 10^{11}(4.59\times 10^{11})$ GeV for $\tilde{g}=8\times 10^{-5}(1.6\times10^{-4})$ in case of $\sigma\phi\chi^2$ interaction. We find that the lower bound of temperature is indeed approximately the same for different coupling and interaction as represented in Fig.\ref{MSSbound}. 
However, given the intimate relation between chaos and squeezing, can the squeezing of the daughter particle states also encode the information about the temperature of the system under consideration?
In the following we conjecture that squeezed state indeed can encode the temperature of the system under consideration as follows:  
   
{\bf Temperature from Squeezed State:} We have already obtained the squeezed state by the application of unitary evolution operator $\hat{\mathcal{U}}_k$ on an unsqueezed two-mode vacuum in the Eq. (\ref{3.27}). For that squeezed state we can define the density matrix for two modes as
\begin{equation}\label{densitymat}
    \hat{\rho}(\vec{k},-\vec{k})= \frac{1}{\cosh^2 r_k}\sum_{n,n^{\prime}=0}^{\infty}e^{2i(n-n^{\prime})\varphi_k}(-1)^{n+n^{\prime}}\tanh^{n+n^{\prime}}r_k\big|n_{\vec{k}},n_{-\vec{k}}\big\rangle\big\langle n^{\prime}_{\vec{k}},n^{\prime}_{-\vec{k}}\big|
\end{equation}
Since one of the two modes is always inaccessible to the observer, we can calculate the reduced density matrix $\hat{\rho}(\vec{k})$ from the full one by tracing out the mode $-\vec{k}$ and  get
\begin{equation}\label{reduceddenmat}
    \hat{\rho}(\vec{k})=\sum_{n=0}^{\infty}\langle n_{-\vec{k}}|\hat{\rho}(\vec{k},-\vec{k})|n_{-\vec{k}}\rangle=\frac{1}{cosh^2r_k}\sum_{n=0}^{\infty}\tanh^{2n} r_k|n_{\vec{k}}\rangle\langle n_{\vec{k}}| \end{equation}
We assume the produced particles are instantaneously thermalized. The above expression of a quantum state can be identified as a thermal squeezed state with inverse temperature defined as \cite{Martin:2015qta}
\begin{equation}\label{inversetemp1}
{\beta}_k=- \ln \tanh^2 r_k .
\end{equation}
We can therefore identify this as a system temperature under consideration.  
As the squeezing parameter $r_k$ is a function of time, $\bar{\beta}_k$ will also vary with time. The very nature of $r_k$ shows a prominent peak near the saturation time scale as seen for $n=1$ model observed in Fig.\ref{sqzparaneq1qua} and Fig.\ref{sqzparaneq123tri}. The temperature of the system being tied with $r_k$ will be maximum at an instant, and for different $k$, this temperature peak position will be different. Collecting only those maximum values, and following the definition of effective MSS bound stated in Eq.\ref{MSSbound2}, we define the average system temperature defined for the thermal squeezed state (SS) over resonance band as
\bea\label{inversetemp2}
\bar{\beta}_{\rm SS} = \frac {1}{{\bar T}_{\rm SS}} =- \frac {2 \int_{\rm band} \ln (\tanh r_k) k^2 dk}{\int_{\rm band}  k^2 dk} .
\eea
We assumed Boltzmann constant $k_B = 1$ in natural unit. Our motive is to check the consistency of the variation of this average temperature for different coupling with the MSS bound and we anticipate that this temperature variation will abide by the MSS bound for any coupling strength. In the absence of any kind of self-interaction or back-reaction, as late time saturation of $r_k$ is obvious in $n=1$ model for both the interaction,  we concentrate only on $n=1$ case in order to define an effective temperature of the system. Using the relation (\ref{inversetemp2}) for two different interactions, we determine an approximate effective temperature that the system can achieve after saturation for different inflaton coupling. For example 
$\bar{T}_{\rm SS} = 6.75 \times 10^{14}$ GeV for $g=2\times 10^{-4}$ in case of $ \phi^2\chi^2$ interaction and $\bar{T}_{\rm SS} = 4\times 10^{13}$ GeV for $\tilde{g}=3\times 10^{-5}$ in case of $\phi\chi^2$ interaction, which are greater than the MSS bound we calculated before. We indeed found the MSS bound to be satisfied for different inflaton coupling and parameters as depicted in Fig.\ref{MSSbound}. To this end, we should remind the reader that the temperature defined above is not the reheating temperature that is defined at the end of reheating.   

 {\bf Rayleigh-Jeans Temperature:} We would like to stress another consideration in the context of thermalization of a system having a very large occupation number. In this limit when the system is in thermal equilibrium the Rayleigh-Jeans spectrum can be defined as $n_kE_k\simeq T_k$, for occupation number $n_k$ and associated energy $E_k$ corresponding to a mode $k$ mode \cite{Podolsky:2005bw,Maity:2018qhi}. In this equation, $T$ is again the temperature of the system under consideration. In our present context energy per mode can be calculated to be $E_k=\bra{1_k}\hat{H}_k\ket{1_k}= \Big(\frac{\Omega_k^2}{2\omega_k}+\frac{\omega_k}{2}\Big)$ using the Hamiltonian $H_X$ of the produced massless fluctuation in Eq.(\ref{3.11}). As the occupation number is usually very high during the preheating phase, to have a better understanding of the thermalization across all the resonant modes for a particular coupling, we can use the combination $n_kE_k$. Following the Eq.(\ref{inversetemp2}) here also we can define an average temperature using the Rayleigh-Jeans(RJ) formula as
\begin{equation}\label{Rayleightemp}
  {\bar T}_{\rm RJ}=\frac{\int_{{\rm band}}n_kE_kk^2dk}{\int_{{\rm band}}k^2dk}  .
\end{equation}
Most interestingly, we find that the equilibrium temperature predicted by the thermal squeezed state in Eq.(\ref{inversetemp2}) matches with the thermalized temperature of the system predicted by the Rayleigh-Jeans formula given in Eq.(\ref{Rayleightemp}) pretty well as shown in Fig.\ref{MSSbound} by red and green dots accordingly. This indeed establishes a beautiful connection between the non-equilibrium squeezed quantum system with the non-equilibrium preheating phase in the early universe. 

Since our present analysis is far from complete, and hence the estimation of different temperatures may not strictly be taken as actual system temperature. At this juncture, therefore, it would be worth pointing out the standard expectation, and we would like to close this section by indicating a naive estimation of the maximum possible temperature the system may attain right after the end inflation, and that can be calculated by utilizing the following relation $H^2_{\rm end} = 1/(3M_{\rm pl}^2) \rho = \pi^2 g/(90 M_{\rm pl}^2) T^4$ with $g\sim 107$ being the effective number of degrees of freedom. In the expression, we considered Stephan-Boltzmann law temperature of radiation energy density $\rho =(g \pi^2/30) T^4$. Assuming $H_{\rm end} \sim 4\times 10^{-6} M_{\rm pl}$, one gets $T \sim 2.6 \times 10^{15}$ GeV which anyway falls within the estimation displayed in Fig.\ref{MSSbound}. The over and under estimation of the temperature with respect to the perturbative temperature just mentioned above can be evaded by the full lattice simulation which we consider for our future study. 

\section{Conclusions}\label{sec13}
Understanding the non-perturbative dynamics of particle production is of immense importance, particularly in the quantum field theory framework. In the realm of cosmology, the early universe reheating phase is an interesting laboratory to understand such phenomena. During the reheating phase, inflaton plays the role of periodic driving force which may lead to resonant particle production if the appropriate conditions are satisfied. Extensive exploration has been performed to understand such non-perturbative phenomena with the motivation to successfully reheat our universe. In this submission, we re-explored such non-perturbative production in the preheating stage and analyzed its underlying physics in relation to quantum squeezing, chaos, and their intriguing connection with thermalization. To the best of our knowledge along that direction very little attention has been paid (see early references \cite{Son:1996uv,McDonald:1999hd,Harigaya:2013vwa,Mukaida:2015ria}) mainly because of a lack of understanding of the possible connection between the reheating and its subsequent evolution to standard cosmology. However, preheating can be assumed as an ideal phase of non-equilibrium processes which demand very special attention from the point of view of theoretical understanding of early universe cosmology. The mechanism of thermalization is of great theoretical importance on its own which is not very well understood. Particularly during reheating, it is believed to have played a significant role in some well-studied cosmological phenomena such as producing matter-anti-matter asymmetry (baryogenesis), and dark matter abundance that we observe today. In the context of perturbative reheating and Lattice studies, the issue of thermalization has been discussed \cite{Podolsky:2005bw,Felder:2000hr,Micha:2004bv,Maity:2018qhi}. However, a theoretical understanding of this is still far from complete. In this paper for the first time we initiate a theoretical exploration of the phenomena of squeezing, chaos, and their deep connection with thermalization during preheating, and we advocate through this submission that we have some promising results that are worth exploring with greater detail in the future. 
For our present study, we have considered two well-studied inflaton interactions with the daughter field namely $\phi\chi^2$ and $\phi^2\chi^2$. After the inflation, the inflaton field oscillates near the minimum of its potential $\phi^n$ and acts as a quasi-periodic driving force leading to chaotic growth of quantum fluctuation which is observed to be imprinted in the OTOC of phase space variables. We have computed the associated effective Lyapunov exponent $(\tilde{\lambda}^{\rm eff}_k)$ characterizing the chaos in the system. During this chaotic growth, the quantum state of the produced particles evolved into a squeezed state. We have further explored the underlying connection between the chaotic growth and the squeezing of the quantum states and showed the connection between Lyapunov exponent and squeezing parameters. Thermalization is believed to be deeply connected to the chaotic behavior of a system, which is beautifully conjectured \cite{Maldacena:2015waa} by proposing an inequality relating Lyapunov exponent and system temperature under consideration $\lambda \gtrsim T$. By using this we calculated the approximate lower bound of temperature ${\bar T}_{\rm MSS}$. We further conjectured a relation between the system temperature and quantum squeezing averaged over phase space and calculated the temperature ${\bar T}_{\rm SS}$ (see Eq.\ref{inversetemp2}) which indeed satisfies the MSS bound ${\bar T}_{\rm SS} > {\bar T}_{\rm MSS}$. Finally to validate our aforesaid conjecture we resort to Rayleigh-Jeans definition of the temperature of a system being in thermal equilibrium. In the classical limit when the occupation number is very large, a thermalized system satisfies the well-known Rayleigh-Jeans formula $nE \sim T$ with $(n, E)$ being the occupation number and particle energy accordingly. Surprisingly, the temperature ${\bar T}_{\rm RJ}$ derived from the Rayleigh-Jeans formula turned out to be ${\bar T}_{\rm RJ} \approx {\bar T}_{\rm SS}$, which is quite consistent with our definition. To this end, we find that among three different background models, $n=1,2,3$, only $n=1$ appears to have a close resemblance to the thermalization process of a typical thermodynamic system under perturbation. The process of thermalization itself is a complex phenomenon. In particular, how a quantum system initially prepared in far-from-equilibrium states can evolve into a thermal equilibrium state is still not a completely understood subject. Eigenstate Thermalization Hypothesis \cite{Srednicki:1994mfb,DAlessio:2015qtq,Foini:2018sdb,Brenes:2021bjr} is a well-known proposal towards this direction. 
In this context, it would be interesting to investigate this ETH  in the reheating era which we have left for our future endeavor.

\section{acknowledgments}
AC would like to thank the Ministry of Human Resource Development, Government of India (GoI), for financial assistance. DM wishes to acknowledge support from the Science and Engineering Research Board~(SERB), Department of Science and Technology~(DST), Government of India~(GoI), through the Core Research Grant CRG/2020/003664.

\bibliographystyle{apsrev4-1}
\bibliography{AYANreferences}

\end{document}